\documentstyle[12pt,epsfig]{article}

\newlength{\dinwidth}
\newlength{\dinmargin}
\def\gsim{\:\raisebox{-0.5ex}{$\stackrel{\textstyle>}{\sim}$}\:}
  {\begin{list}{}{\settowidth{\labelwidth}{#1}%
  \setlength{\leftmargin}{\labelwidth}%
  \addtolength{\leftmargin}{\labelsep}%
  \setlength{\itemsep}{0pt plus 1pt}
  \setlength{\parsep}{0pt plus 1pt}
  \setlength{\topsep}{0pt plus 1pt}
  \setlength{\partopsep}{0pt plus 1pt}
  \setlength{\parskip}{2mm plus 1mm minus 1mm}
  }}%
  {\end{list}}
\begin{document}
\newcommand{\vs}{\vspace{5mm}}
\newcommand{\ecms}{\sqrt{s}}
\newcommand{\lattice}{
\multiput(0,0)(10,0){18}{\line(0,1){200}}
\multiput(0,0)(0,10){21}{\line(1,0){170}}
}
%
\begin{flushright}
  BI-TP 97/34\\ WUE-ITP-97-034\\ MPI-PhE/97-23\\[1.7ex] 
  {\tt hep-ph/9709315} \\ 
\end{flushright}

\vskip 35pt

\begin{center}
\begin{Large}
\begin{bf}
  Leptoquark Pair Production \\ 
  at $e^+e^-$ Linear Colliders: \\ 
  Signals and Background \\ 
\end{bf}
\end{Large}

\vspace{13mm}

{\large R.\ R\"uckl$^{1,2}$, R.\ Settles$^2$ and H.\ Spiesberger$^3$}
\\ 

\vspace{10mm}

{\em $^1$ Institut f\"ur Theoretische Physik, Universit\"at W\"urzburg,\\
D-97074 W\"urzburg, Germany}\\[1.1ex]
{\em $^2$ Max-Planck-Institut f\"ur Physik, Werner-Heisenberg-Institut,\\
D--80805 M\"unchen, Germany}\\[1.1ex]
{\em $^4$ Fakult\"at f\"ur Physik, Universit\"at Bielefeld,\\
D-33501 Bielefeld, Germany}\\[2ex]

\vspace{10mm}

\today

\vspace{20mm} 

{\bf ABSTRACT}

\end{center}

\begin{quotation}
  We study pair production of leptoquarks in $e^+e^-$ annihilation at
  linear colliders. A detailed simulation including beamstrahlung and
  initial state radiation, leptoquark decay and hadronization, as well
  as detector smearing, is performed. Discovery limits are estimated
  for center-of-mass energies of 500 and 800 GeV. The prospects for
  determining masses and couplings of leptoquarks are also
  investigated.
\end{quotation}

\vfill
\footnoterule
{\footnotesize
\noindent
To appear in the Proceedings of the {\em Joint ECFA/DESY Study on
  Physics and Detectors for the Linear Collider}, February to November
  1996, ed.\ R.\ Settles, DESY 97-123E.}

\newpage
\section{Introduction}

Leptoquarks appear in extensions of the standard model involving
unification, technicolor, compositeness, or $R$-parity violating
supersymmetry. Hence, the search for leptoquarks is an important task
at present and future high energy experiments. Furthermore,
leptoquark pair production leads to novel experimental signatures
involving fixed-mass lepton-jet systems, and thus provides an
interesting example for detector studies.

In $e^+e^-$ annihilation, pair production of scalar leptoquarks by
$s$-channel $\gamma$ and $Z$ exchange is uniquely determined by gauge
symmetry if one assumes minimal couplings, while vector leptoquarks
may also possess anomalous couplings. In addition to $s$-channel
production, there are $t$-channel processes involving unknown Yukawa
couplings. Similarly, pair production of leptoquarks at hadron
colliders is determined by the QCD gauge couplings with contributions
from anomalous and Yukawa couplings. In contrast, in lepton-nucleon
scattering the dominant $s$-channel leptoquark production, as well as
$t$- and $u$-channel scattering, proceeds via Yukawa couplings
only.  

In the generally adopted framework described in Ref.\ \cite{RB}, the
Yukawa couplings are taken to be dimensionless and $SU(3)\times
SU(2)\times U(1)$ symmetric. Moreover, they are assumed to conserve
lepton- and baryon-number in order to avoid rapid proton decay, to be
non-zero only within one family in order to exclude FCNC processes
beyond the CKM mixing, and chiral in order to avoid the very strong
bounds from leptonic pion decays.  The allowed states can be
classified according to spin, weak isospin and fermion number. They
are summarized in Table \ref{tabprop}.

The leptoquark masses and couplings are constrained by high-energy data.
Direct searches for leptoquarks have been performed at the Tevatron, at
HERA and at LEP. Recently, both experiments, CDF and D0, have improved
their mass limits for scalar leptoquarks considerably.  D0 excludes
first generation leptoquarks with masses below 225 GeV assuming a
branching ratio $B_{eq}=1$ for decays into electrons \cite{D0}, whereas
CDF quotes a limit of 213 GeV \cite{cdf} (all mass limits are at 95\,\%
CL). For branching ratios less than one, the limits are weaker, e.g., $M
> 176$ GeV for $B_{eq}=0.5$ \cite{D0}. The bounds on vector states are
even stronger: 298 GeV for $B_{eq}=1$ and 270 GeV for $B_{eq}=0.5$
\cite{vector}. The corresponding bounds on second and third generation
scalar leptoquarks are $M > 184$ GeV for $B_{\mu q} = 1$ and $M > 98$
GeV for $B_{\tau q} = 1$, respectively \cite{lqst}. The LHC is expected
to reach the TeV range. Mass limits obtained at HERA depend on the
Yukawa couplings $\lambda_{L,R}$. They range from 207 to 272 GeV for
$\lambda_{L,R} = e$ \cite{hera} where $e$ is the electromagnetic
coupling strength .  The precise mass limit also varies with the type of
leptoquark specified in Table \ref{tabprop}.  The above limits are
lowered by about 50 GeV if $\lambda_{L,R} =0.1$. Finally, the most
stringent but $\lambda_{L,R}$-dependent limit from LEP originates from
the search for single-leptoquark production and excludes masses below
131 GeV assuming $\lambda_{L,R} > e$ \cite{leps}. The mass bounds from
leptoquark pair production roughly reach $\sqrt{s}/2$ \cite{lepp},
$\sqrt{s}$ being the center-of-mass energy, and are thus much weaker
than the Tevatron bounds.

Indirect bounds on Yukawa couplings and masses can be derived from
low-energy data \cite{leurer}. For chiral couplings, the most
restrictive limits come from atomic parity violation and lepton and
quark universality.  The maximum allowed couplings for $M=200$ GeV and
first generation leptoquarks are given in Table \ref{tabprop}
\cite{krsz}.

\begin{table}[hbp]
\begin{center}
\begin{tabular}{|c|c|c|c|c|c|r|r|}
\hline \rule{0mm}{5mm}
  & State & ${}^Q\bar{\Phi}_T$              &
$\lambda_{L,R}$ & Channel & Limits & $g_1(s)$ & 
$\sigma_{tot}(s) [$fb$]$ \\[1mm]
\hline \rule{0mm}{5mm}
1 & $S_1$ & ${}^{-1/3}S_0$ & 
$\displaystyle \begin{array}{c} g_L \\ g_R \\ -g_L \end{array}$ &  
$\displaystyle \begin{array}{c} e_L^- u_L \\
                            e_R^- u_R \\ \nu_e d_L \end{array}$ &
   {\raisebox{-0.8ex}{$\displaystyle 
   \begin{array}{c} g_L<0.06 \\ 
       \rule{0mm}{6mm} {\raisebox{1ex}{$g_R<0.1$}} \end{array}$}} &
                                           $0.236~$ & $6.15~$ \\[1mm]
\hline \rule{0mm}{5mm}
2 & $\tilde{S}_1$ & ${}^{-4/3}\tilde{S}_0$      &
$g_R$ & $e_R^- d_R$ & $g_R<0.1$    & $3.77~$ & $98.2~$\\[1mm]
\hline \rule{0mm}{5mm}
  &               & ${}^{+2/3}S_1$              &
$\sqrt{2}g_L$  & $\nu_e u_L$ &  &
$4.23~$ & $110~$ \\
3 & $S_3$         & ${}^{-1/3}S_1$              &
$\displaystyle \begin{array}{c} -g_L \\ -g_L \end{array}$         
& $\displaystyle \begin{array}{c}
                 \nu_e d_L \\       
                 e_L^- u_L \end{array}$ &  $g_L<0.09$
                                      & $0.236~$ & $6.15~$ \\
  &               & ${}^{-4/3}S_1$              &
$-\sqrt{2}g_L$ & $e_L^- d_L$ &  & $6.05~$ & $158~$ \\[1mm]
\hline \rule{0mm}{5mm}
4 & $R_2$ & 
$\displaystyle \begin{array}{c} {}^{-2/3}S_{1/2} \\ \\
                                {}^{-5/3}S_{1/2} \end{array}$ &
$\displaystyle \begin{array}{c} g_L \\ -g_R \\ g_L \\ g_R\end{array}$&
$\displaystyle \begin{array}{c} \nu_e \bar{u}_L \\
                                e_R^- \bar{d}_R \\
                                e_L^- \bar{u}_L \\
                                e_R^- \bar{u}_R \end{array}$ &
$\displaystyle \begin{array}{r} g_L<0.1 \\ \\ g_R<0.09 \end{array}$ &
$\displaystyle \begin{array}{r} 2.52 \\ \\ 5.70 \end{array}$ &
$\displaystyle \begin{array}{r} 65.8 \\ \\ 149 \end{array}$ \\[6mm]
\hline \rule{0mm}{8mm}
5 & $\tilde{R}_2$ & $\displaystyle \begin{array}{c} 
                  {}^{+1/3}\tilde{S}_{1/2} \\
                  {}^{-2/3}\tilde{S}_{1/2} \end{array}$ &
$\displaystyle \begin{array}{c} g_L \\ g_L \end{array}$ & 
$\displaystyle \begin{array}{c} \nu_e \bar{d}_L \\
                                e_L^- \bar{d}_L \end{array}$ & 
$g_L < 0.1$ &
$\displaystyle \begin{array}{r} 1.06 \\ 1.51 \end{array}$ & 
$\displaystyle \begin{array}{r} 27.6 \\ 39.5 \end{array}$ \\[4mm]
\hline
\hline \rule{0mm}{5mm}
6 & $V_2$ & 
$\displaystyle \begin{array}{c} {}^{-1/3}V_{1/2} \\ \\
                                {}^{-4/3}V_{1/2} \end{array}$ &
$\displaystyle \begin{array}{c} g_L \\ g_R \\ g_L \\ g_R\end{array}$&
$\displaystyle \begin{array}{c} \nu_e d_R \\
                                e_R^- u_L \\
                                e_L^- d_R \\
                                e_R^- d_L \end{array}$ &
$\displaystyle \begin{array}{r} g_L<0.09 \\ \\ g_R<0.05 \end{array}$ &
$\displaystyle \begin{array}{r} 1.56 \\ \\ 3.84 \end{array}$ &
$\displaystyle \begin{array}{r} 365 \\ \\ 895 \end{array}$ \\[6mm]
\hline \rule{0mm}{8mm}
7 & $\tilde{V}_2$ & $\displaystyle \begin{array}{c} 
                  {}^{+2/3}\tilde{V}_{1/2} \\
                  {}^{-1/3}\tilde{V}_{1/2} \end{array}$ &
$\displaystyle \begin{array}{c} g_L \\ g_L \end{array}$ & 
$\displaystyle \begin{array}{c} \nu_e u_R \\
                                e_L^- u_R \end{array}$ & 
$g_L < 0.09$ &
$\displaystyle \begin{array}{r} 1.51 \\ 1.06 \end{array}$ & 
$\displaystyle \begin{array}{r} 353 \\ 247 \end{array}$ \\[4mm]
\hline \rule{0mm}{5mm}
  &   &   & $g_L$ & $e_L^- \bar{d}_R$ & 
  {\raisebox{-1ex}{$g_L < 0.05$}} &  & \\
8 & $U_1$ & ${}^{-2/3}V_{0}$ & $g_R$ & $e_R^- \bar{d}_L$ &  
                                              & $0.942~$ & $222~$\\
  &   &   & $g_L$ & $\nu_e \bar{u}_R$ & 
  {\raisebox{1ex}{$g_R < 0.09$}} &   & \\[1mm]
\hline \rule{0mm}{5mm}
9 & $\tilde{U}_1$ & ${}^{-5/3}\tilde{V}_0$      &
$g_R$ & $e_R^- \bar{u}_L$ & $g_R<0.09$& $5.89~$ & $1370~$\\[1mm]
\hline \rule{0mm}{5mm}
  &               & ${}^{+1/3}V_1$              &
$\sqrt{2}g_L$  & $\nu_e \bar{d}_R$ &  & $4.02~$ & $942~$ \\
10 & $U_3$         & ${}^{-2/3}V_1$              &
$\displaystyle \begin{array}{c} -g_L \\ g_L \end{array}$         
& $\displaystyle \begin{array}{c}
                 e_L^- \bar{d}_R \\
                 \nu_e \bar{u}_R \end{array}$ & $g_L < 0.04$ 
                                      & $0.942~$ & $222~$ \\
  &               & ${}^{-5/3}V_1$              &
$\sqrt{2}g_L$ & $e_L^- \bar{u}_R$ &  & $7.67~$ & $1790~$ \\[1mm]
\hline
\end{tabular}
\caption{\it Properties of leptoquarks. Columns 2 and 3 show the
  notations of Refs.\ \protect\cite{RB} and \protect\cite{koehler},
  respectively. The upper half of the Table refers to scalars, the
  lower to vectors. The lower indices denote the weak isospin,
  multiplicity $2T+1$ or total isospin $T$. The upper index is the
  electromagnetic charge. The Yukawa couplings $g_{L,R}$ are as
  defined in Ref.\ \protect\cite{RB}, while $\lambda_{L,R}$ is used as
  a generic symbol. Column 5 specifies the production and decay
  channels, while the low-energy limits on the Yukawa couplings from
  Ref.\ \protect\cite{leurer} for $M=200$ GeV are given in column 6.
  The last two columns quantify the effective coupling $g_1$(s)
  defined in Eq.\ (\ref{coupg1}) and the total cross section
  $\sigma_{tot}(s)$ given in Eqs.\ (\ref{sigtot}, \ref{sigtotf}) taking
  $\protect\sqrt{s} = 500$ GeV, $M=200$ GeV, $g_L=g_R=0$, and
  including corrections due to beamstrahlung and initial state
  radiation.}
\label{tabprop}
\end{center}
\end{table}

The present work extends previous studies \cite{epem} in several ways:
\begin{itemize}
\item[(i)] complete simulation of leptoquark production and decay
  \begin{equation}
  e^+e^- \rightarrow \Phi\bar{\Phi}
         \rightarrow l_1q_1 \bar{l}_2\bar{q}_2
         ~~~{\rm or} ~~~ l_1\bar{q}_1 \bar{l}_2 q_2,
  \end{equation}
  taking into account beamstrahlung, initial-state radiation and quark
  hadronization;
\item[(ii)] simulation of the main background processes
  \begin{equation}
  e^+e^- \rightarrow WW, ~~ ZZ,
  \end{equation}
  \begin{equation}
  e^+e^- \rightarrow l_1\bar{l_2} q_3\bar{q_4},
  \end{equation}
  \begin{equation}
  e^+e^- \rightarrow \bar{t}t;
  \end{equation}
\item[(iii)] inclusion of acceptance and smearing effects for different
  detector models.
\end{itemize}
After the presentation of calculational details in section 2 and
experimental considerations in section 3, the results on detection
efficiencies and sensitivity limits are discussed in section 4.

\section{Theoretical Framework}

In $e^+e^-$ annihilation, leptoquarks can be pair-produced via photon
and $Z$ boson $s$-channel exchange and via $t$-channel exchange of
quarks. The couplings of leptoquarks to gauge bosons are given in Ref.\ 
\cite{BluRu}\footnote{For a generalization including anomalous couplings
  of vector states, see Ref.\ \cite{bbk}. The couplings to gauge bosons
  used here correspond to minimal couplings with $\kappa = 1$ and
  $\lambda = 0$.}, the couplings to fermions in Ref.\ \cite{RB}.  With
these couplings, the differential cross section for the production of
scalar or vector leptoquarks of mass $M$ is given by \cite{BluRu}
\begin{equation}
\frac{d\hat{\sigma}}{d\cos\theta} = 
\frac{3\pi\alpha^2}{8} N \left\{ \sin^2\theta 
\left[ A_0 + \frac{A_1}{t} + \frac{A_2}{t^2} \right]
+ B_0 + \frac{B_1}{t} \right\},
\label{sigdiff}
\end{equation}
where $t = 1+\beta^2 - 2\beta\cos\theta$ with $\beta = \sqrt{1 -
  4M^2/s}$.  The coefficients $N$, $A_i$ and $B_i$ are listed in Table
\ref{sigcoeff}, the effective coupling parameters $g_i$ being defined
by
\begin{equation}
g_1 = \sum_{a=L,R} |\kappa_a(s)|^2
\label{coupg1}
\end{equation}
\begin{equation}
g_2 = 4 \sum_{a=L,R} \left(\frac{\lambda_a}{e}\right)^2 {\rm
  Re}\kappa_a(s)
\label{coupg2}
\end{equation}
\begin{equation}
g_3 = 4 \sum_{a=L,R} \left(\frac{\lambda_a}{e}\right)^4
\label{coupg3}
\end{equation}
with
\begin{equation}
\kappa_a(s) = \sum_{V=\gamma,Z} Q_a^V(e)
\frac{s}{s-M_V^2+iM_V\Gamma_V} Q^V(\Phi), 
\end{equation}
and $\lambda_a$ denoting the Yukawa couplings specified in Table
\ref{tabprop}.  In the above, $e=\sqrt{4\pi\alpha}$,
$M_{\gamma}=\Gamma_{\gamma}=0$, and $M_Z$ and $\Gamma_Z$ are the mass
and the width of the $Z$ boson, respectively. The electron and
leptoquark electroweak charges are given by
\begin{equation}
\begin{array}{lll}
Q_{L,R}^{\gamma}(e) = -1, &
\displaystyle
Q_L^Z(e) = \frac{-1/2 + s_w^2}{s_wc_w}, &
\displaystyle
Q_R^Z(e) = \frac{s_w}{c_w}, \\[1em]
\displaystyle
Q^{\gamma}(\Phi) = Q, & \displaystyle
Q^Z(\Phi) = \frac{T_3-Q s_w^2}{c_ws_w} &
\end{array}
\end{equation}
where $Q$ is the electric charge in units of $e$, $T_3$ the third
component of the weak isospin, and $s_w$ ($c_w$) the sine (cosine) of
the weak mixing angle.

\begin{table}[htb]
\begin{center}
\begin{tabular}{|c|c|c|}
\hline \rule{0mm}{5mm}
      & scalar & vector \\[1mm]
\hline \rule{0mm}{8mm}
$N$       & $\displaystyle \frac{\beta^3}{s}$ 
          & $\displaystyle \frac{\beta}{M^2}$ \\[3mm]
\hline \rule{0mm}{8mm}
$A_0$ &  $g_1$    & $\displaystyle
                          \frac{1}{4}\beta^2(1-3\beta^2)g_1
                         -\frac{1}{4}\beta^2g_2
                         +\frac{\beta^2s}{16M^2}g_3$ \\[3mm]
\hline \rule{0mm}{8mm}
$A_1$ &  $g_2$    & $\displaystyle
\frac{1}{2}\beta^2(1-\beta^2)g_2$ \\[3mm]
\hline \rule{0mm}{6mm}
$A_2$ &  $g_3$   & $\displaystyle \beta^2(1-\beta^2)g_3$  \\[1mm]
\hline \rule{0mm}{8mm}
$B_0$ &    0      & $\displaystyle \beta^2g_1
                          + \frac{1}{2}(1+\beta^2)g_2 + g_3$  \\[3mm]
\hline \rule{0mm}{8mm}
$B_1$ &    0      & $\displaystyle -\frac{1}{2}(1-\beta^2)^2g_2$
\\[3mm]
\hline
\end{tabular}
\caption{\it Coefficients appearing in the cross section formula Eq.\ 
  (\ref{sigdiff}).}
\label{sigcoeff}
\end{center}
\end{table}

The total cross section can be written in the following form
\cite{BluRu,bbk}:
\begin{equation}
\hat{\sigma} = \frac{\pi\alpha^2}{2} N
\left( F_1^J g_1 + F_2^J g_2 + F_3^J g_3 \right)
\label{sigtot}
\end{equation}
using the same effective couplings $g_i$, $i=1$, 2, 3 defined in Eqs.\ 
(\ref{coupg1}, \ref{coupg2}, \ref{coupg3}) and
\begin{equation}
\begin{array}{l}
\displaystyle
F_1^S = 1, 
\\[1ex]
\displaystyle
F_2^S = \frac{3}{8} \left( \frac{1+\beta^2}{\beta^2} -
  \frac{(1-\beta^2)^2}{2\beta^3} \ln \frac{1+\beta}{1-\beta} \right), 
\\[2.5ex]
\displaystyle
F_3^S = \frac{3}{4} \left( -\frac{1}{\beta^2} +
  \frac{1+\beta^2}{2\beta^3} \ln \frac{1+\beta}{1-\beta} \right),
\\[3.5ex]
\displaystyle
F_1^V = \beta^2\frac{7-3\beta^2}{4}, 
\\[2ex]
\displaystyle
F_2^V = \frac{15}{16} + \frac{1}{2}\beta^2 - \frac{3}{16}\beta^4
   -\frac{3}{32\beta} \left(1-\beta^2\right)^2 \left(5-\beta^2\right)
   \ln \frac{1+\beta}{1-\beta}, 
\\[2ex]
\displaystyle
F_3^V = \frac{3}{4} \left(1+\beta^2\right) +
\frac{\beta^2}{16}\frac{s}{M^2} 
\frac{3}{8\beta}\left(1-\beta^4\right) 
\ln \frac{1+\beta}{1-\beta}.
\end{array}
\label{sigtotf}
\end{equation}
The normalization factor $N$ is as given in Table \ref{sigcoeff}.
Asymptotically, for $M^2 \ll s$, the $s$-channel and the
$s/t$-interference contributions scale like $1/s$ for scalar leptoquarks
and approach constant values for vectors. The pure $t$-channel
contribution scales with $\frac{1}{s} \ln (s/M^2)$ for scalars, but
grows like $s/M^4$ for vectors. The latter behavior eventually leads to
unitarity violation and thus indicates that the effective Lagrangian
from which the above cross section formulae have been derived has to be
embedded into a more fundamental theory at high energies. However, for
Yukawa couplings obeying the experimental bounds of Table \ref{tabprop}
this problem only arises far beyond the energy range considered here.

At high-luminosity $e^+e^-$ colliders, beamstrahlung leads to a
significant energy spread in the electron and positron beams.  This
effect can be described with the help of a radiator function
$G_{e^+e^-}(x_+,x_-,s)$ where $x_{\pm}$ is related to the fractional
energy loss, $E_{\pm} = x_{\pm}\sqrt{s}/2$.  The transverse momentum
spread of the beam due to beamstrahlung can be neglected.  Therefore,
the cross sections including the effect of beamstrahlung can be
obtained from Eqs.\ (\ref{sigdiff}) and (\ref{sigtot}) by a simple
convolution:
\begin{equation}
d\sigma_{bs}(s) = \int dx_+ dx_- d\hat{\sigma}(x_+x_-s)
G_{e^+e^-}(x_+,x_-,s).
\label{dsigmabs}
\end{equation}
Because of singularities at the end-points $x_+ = 1$ and $x_- = 1$,
$G_{e^+e^-}(x_+,x_-)$ is split into $\delta$-function terms (the
probability for no energy loss by beamstrahlung) and a smooth function
of $x_+$ and $x_-$.  The radiator function depends on machine
parameters and is numerically provided by the program {\tt circe}
\cite{circe}. 

\begin{figure}[htb] 
\unitlength 1mm
\begin{picture}(160,88)
\put(30,-25){
\epsfxsize=10cm
\epsfysize=14cm
\epsfbox{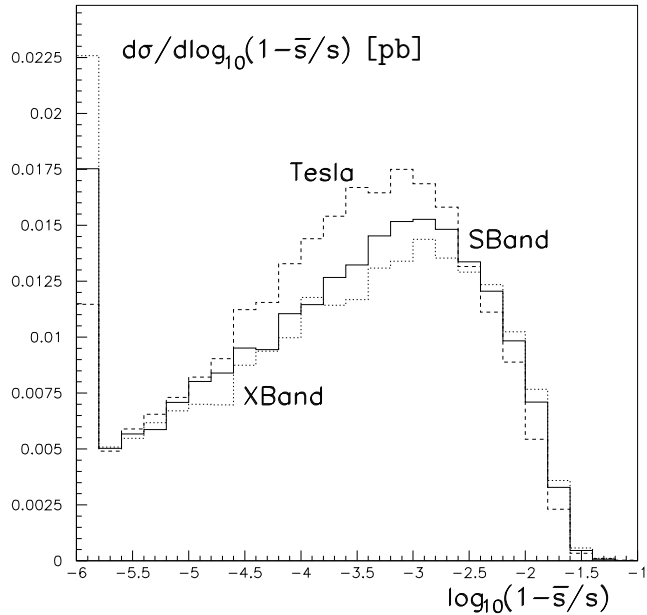}
}
\put(82,75){{\tt [pb]}}
\end{picture}
\caption{\it Distribution of the production cross section in the
  effective collision energy $\protect\sqrt{\bar{s}}$ for the scalar
  leptoquark ${}^{-4/3}S_1$. The three histograms refer to the Tesla
  (dashed), S-Band (full) and X-Band (dotted) options for
  $\protect\sqrt{s} =500$ GeV, $M = 200$ GeV, $\lambda_L = \lambda_R =
  0$.}
\label{figbeamstr}
\end{figure}

The most important QED correction arises from initial state radiation
(ISR). The universal part of this effect can again be described by a
radiator function\footnote{$D_{e/e}$ is actually known up to
  $O(\alpha^2)$, but since the non-universal contributions to
  $O(\alpha)$ are not known for the process under consideration, we
  also do not include the $O(\alpha^2)$ effects in $D_{e/e}$.}:
\begin{equation}
D_{e/e}(z) = 
\delta(1-z)\left[1 + \frac{\alpha}{2\pi} \ln\frac{4E^2}{m_e^2} 
\left(2\ln\epsilon + \frac{3}{2}\right)\right]
+ \theta(1-z-\epsilon) \frac{\alpha}{2\pi} \frac{1+z^2}{1-z}
\ln\frac{4E^2}{m_e^2} 
\label{ISRrad}
\end{equation}
where $E$ is the beam energy. The parameter $\epsilon$ is a photon
energy cutoff separating soft from hard photons. The $\delta$-function
term includes soft bremsstrahlung where photons have an energy
$E_{\gamma} < \epsilon E$.  The second term proportional to the
$\theta$-function describes the emission of photons with $E_{\gamma} >
\epsilon E$.  Similarly as in Eq.\ (\ref{dsigmabs}), the corrected cross
section follows by convolution:
\begin{equation}
d\sigma = \int_0^1 dz D_{e/e}(z) d\hat{\sigma}(zE_+,E_-) + 
          \int_0^1 dz D_{e/e}(z) d\hat{\sigma}(E_+,zE_-).
\label{dsigmac}
\end{equation}
In addition, for $s$-channel diagrams, higher-order corrections due to
vacuum polarization can be included by using the running
fine-structure constant \cite{vacpol}.

If one wants to take into account beamstrahlung simultaneously with
ISR, one has to substitute in Eq.\ (\ref{dsigmac}) $d\hat{\sigma}$ by
$d\sigma_{bs}$ given in Eq.\ (\ref{dsigmabs}) and $E_+$ ($E_-$) by
$x_+\sqrt{s}/2$ ($x_-\sqrt{s}/2$). The reduced center-of-mass energy
squared after beamstrahlung and ISR is given by $\bar{s} = x_+x_-zs$.
The separation of $G_{e^+e^-}$ and $D_{e/e}$ into $\delta$-function
terms and smooth functions and the distinction of initial-state
radiation from electrons and positrons requires to compute eight
separate contributions to the cross sections.  Technically, the
mapping $z \rightarrow \ln(1-z)$ leads to numerically stable
integration.

Presently, three different options for the beam dynamics of an
$e^+e^-$ collider are being discussed. The differences in the
distributions of the cross sections in $\bar{s}$ are exemplified in
Fig.\ \ref{figbeamstr} for the leptoquark ${}^{-4/3}S_1$ having the
largest cross section. For the present study, we choose the Tesla
option. The bulk of the events is accompanied by soft radiation with a
maximum at $1-\bar{s}/s \simeq 10^{-3}$, but with a tail extending to
a few percent. This limits the possibility to reconstruct the
kinematics of final states with missing momentum by imposing the
constraints of energy-momentum conservation.

\begin{figure}[htb] 
\unitlength 1mm
\begin{picture}(160,80)
\put(-14,-122){
\epsfxsize=20cm
\epsfysize=24cm
\epsfbox{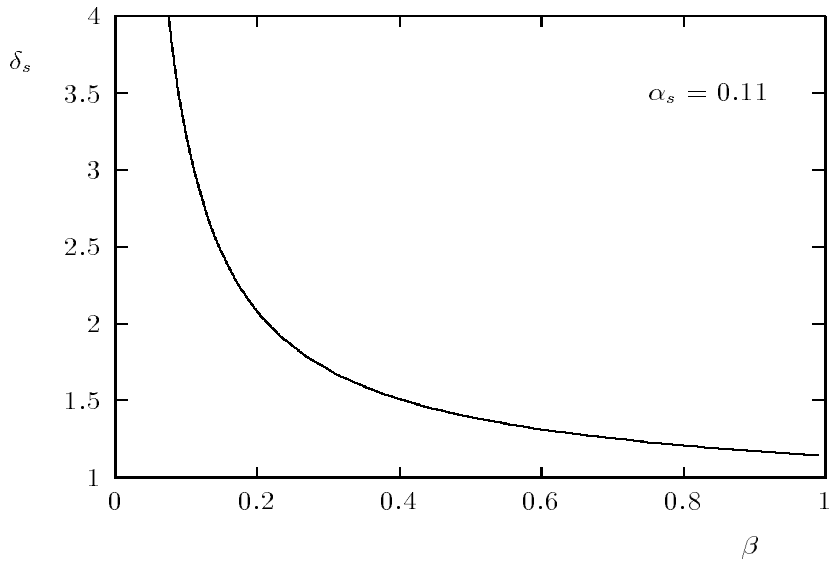}
}
\end{picture}
\caption{\it QCD correction factor $\delta_s$ for the total production
  cross section of scalar leptoquarks as a function of $\beta$.}
\label{figqcdcorr}
\end{figure}

QCD corrections are not taken into account in our calculations. They
are known only for the total cross section of scalar leptoquark
production \cite{qcdcorr}. The correction factor $\delta_s = 1 +
\alpha_s F(\beta)$ is shown in Fig.\ \ref{figqcdcorr}.  At $\beta
\leq 0.2$ the corrections exceed 100$\,\%$. In this region close to
the threshold it will certainly be necessary to take into account
bound state dynamics. At large $\beta$, one expects corrections of the
order of $20\,\%$ which increase the cross section.

\begin{figure}[htb]
\unitlength 1mm
\begin{picture}(160,84)
\put(-16,-48){
\epsfxsize=12.5cm
\epsfysize=15cm
\epsfbox{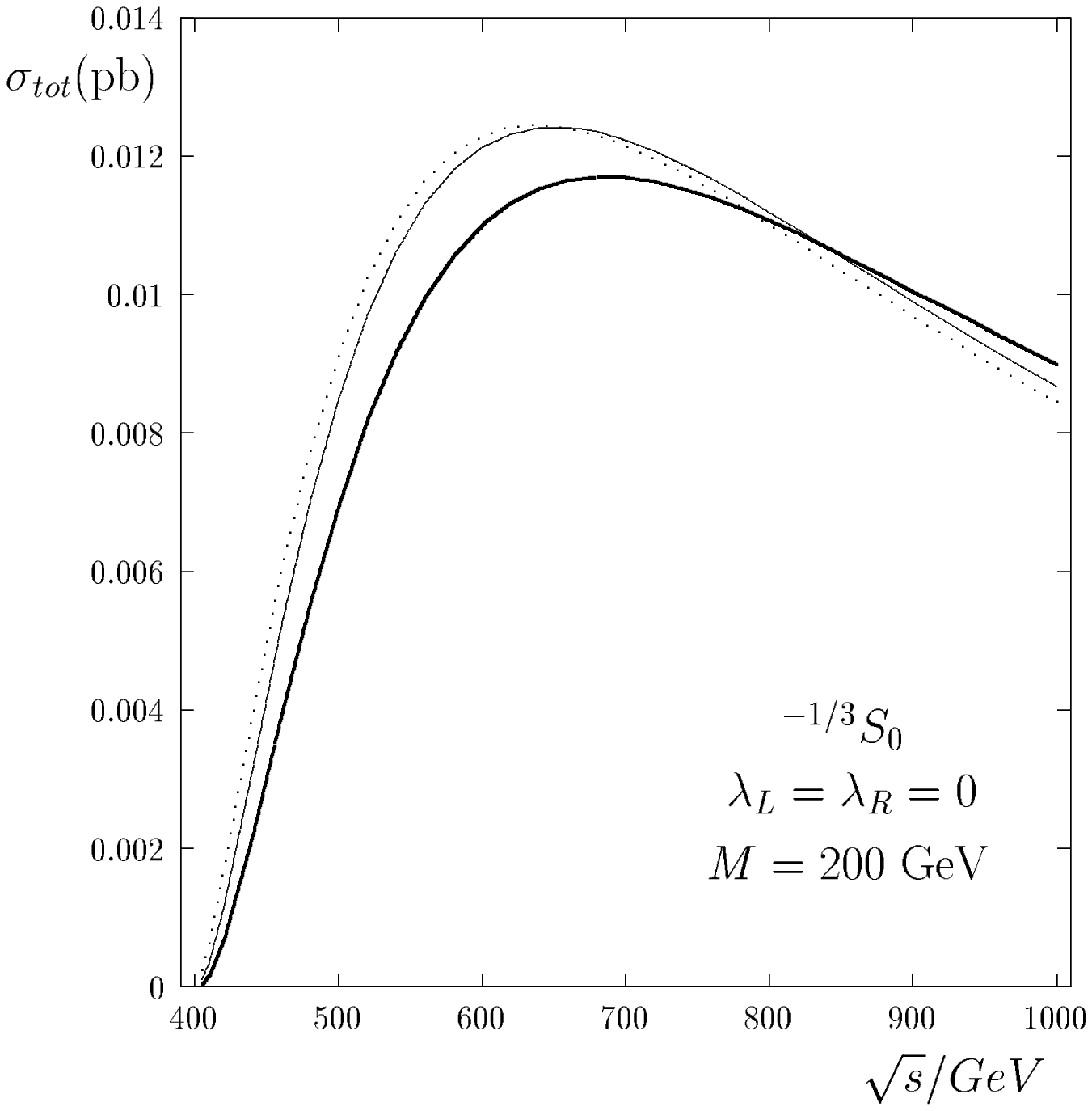}
}
\put(42,0){{\rm (a)}}
\put(65,-48){
\epsfxsize=12.5cm
\epsfysize=15cm
\epsfbox{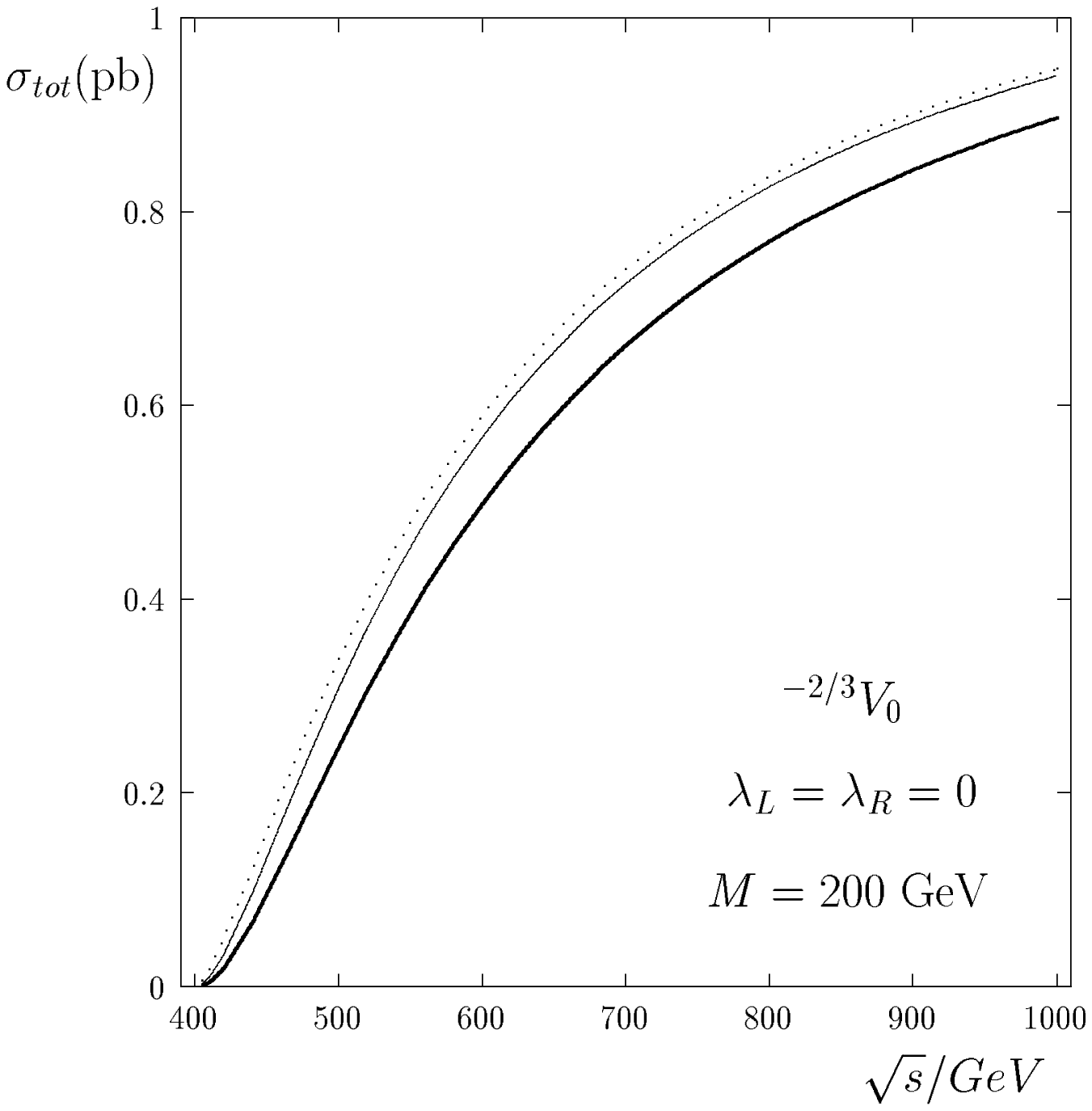}
}
\put(123,0){{\rm (b)}}
\end{picture}
\caption{\it Effects of beamstrahlung and initial state radiation on
  the pair-production cross sections for the ${}^{-1/3}S_0$ (a) and
  ${}^{-2/3}V_0$ (b) leptoquarks.  The dotted curves are the Born
  cross sections, the thin full curves include beamstrahlung, and the
  thick full curves give the total cross section including both
  beamstrahlung and ISR.}
\label{figsigcorr}
\end{figure}

In Fig.\ \ref{figsigcorr}, finally, we illustrate the effects due to 
beamstrahlung and ISR on the integrated cross sections for 
representative scalar and vector leptoquarks. Except for scalars at
very high $\sqrt{s}$, the total cross section is reduced typically by
10\,\%: 1 to 2 fb for scalars and 60 fb for vectors. The reduction is
particularly important for the discovery reach near threshold.

\begin{figure}[htb] 
\unitlength 1mm
\begin{picture}(160,138)
\put(-17,-103){
\epsfbox{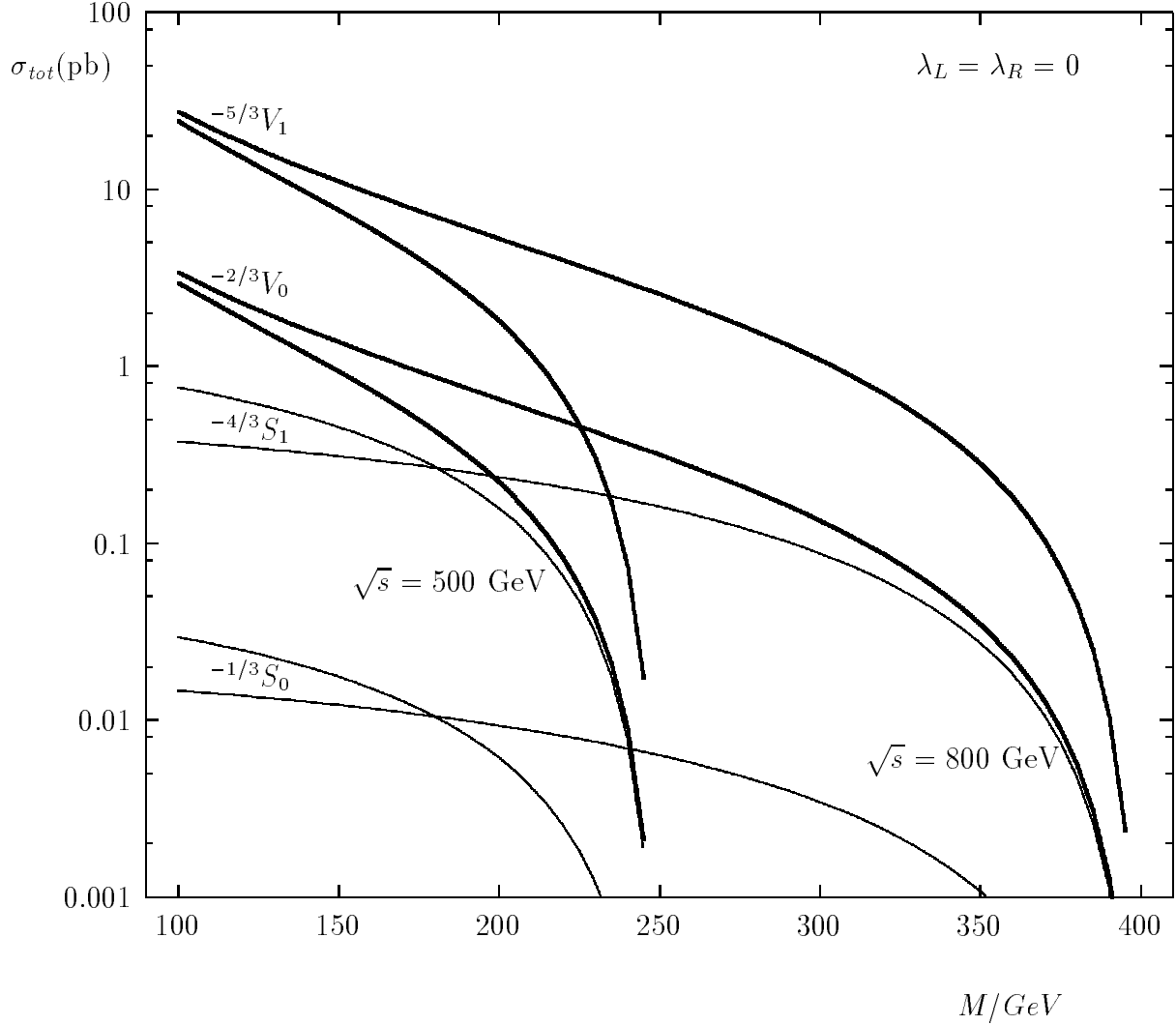}
}
\end{picture}
\caption{\it Total cross sections for leptoquark pair-production at
  fixed center-of-mass energies as a function of the leptoquark mass
  $M$ assuming vanishing Yukawa couplings and including corrections
  due to beamstrahlung and ISR.}
\label{figsigtot}
\end{figure}

Figure \ref{figsigtot} shows the total cross section for leptoquark
pair-production at two center-of-mass energies and for those species
of scalar and vector leptoquarks which have the minimal and maximal
cross sections in each class.  From the figure it is clear that, given
the mass, the measurement of the total cross section already provides
an important piece of information on the type of the leptoquark
produced. As demonstrated later, another important observable is the
angular distribution.

Additional information can be inferred from decay properties. They are
determined by the Yukawa couplings. The partial decay widths per
channel (see Table \ref{tabprop}) are given by \cite{RB}
\begin{equation}
\begin{array}{l}
\displaystyle
\Gamma_S = \frac{1}{16\pi^2} \lambda_{L,R}^2 M ~~~~~~{\rm for~scalars},
\\[1em]
\displaystyle
\Gamma_V = \frac{1}{24\pi^2} \lambda_{L,R}^2 M ~~~~~~{\rm for~vectors}.
\end{array}
\label{widths}
\end{equation}
Quantitatively, for $M=200$ GeV and $\lambda_{L,R} = e$ one has
$\Gamma_S\,\, (\Gamma_V) = 116 \,\, (77)$ MeV, i.e., very narrow states.
Furthermore, for the leptoquarks considered here, the branching ratios
$B_{eq}$ for charged lepton channels and $B_{\nu q}$ for neutrino
channels add up to unity, $B_{eq} + B_{\nu q} = 1$, with $B_{eq}$ as
given in Table \ref{tabbr}.  If one assumes that the members of a given
isomultiplet of leptoquarks are almost mass-degenerate, one sees that
the strongest bounds, $M \gsim 225$ GeV for scalars and $M \gsim 300$
GeV for vectors, obtained at the Tevatron for $B_{eq} = 1$ apply to all
leptoquark species except the two singlets ${}^{-1/3}S_0$ and
${}^{-2/3}V_0$.  The latter two may have branching fractions $1/2 \leq
B_{eq} \leq 1$ so that only the weaker limits $M_{S_0} > 176$ GeV and
$M_{V_0} > 270$ GeV hold. Note that for first generation leptoquarks in
this mass range, $B_{eq}$ must be very close to $1/2$ ($r=0$) or 1
($r=\infty$) unless $\lambda_L \lambda_R \ll 1$.  In more general
scenarios where leptoquarks have additional decay channels, as is the
case for example for squarks in supersymmetric models with $R$-parity
violation, $B_{eq}$ and $B_{\nu q}$ are not fixed by the Yukawa
couplings alone.  They rather can be considered as independent
parameters leading to more model-dependent mass bounds.  We also mention
the possibility for generic leptoquarks to have tiny Yukawa couplings
and hence being very long-lived. In this case, they would have to form
bound states (leptohadrons) which do not decay inside the detector. This
requires special search strategies, e.g., for fractional charged
particles. In our calculations, the branching ratios are always taken
for generic leptoquarks in accordance with the Yukawa couplings chosen
in the production process. Of course, sufficiently small $\lambda_{L,R}$
has negligible influence on the production rates.

\begin{table}[htbp]
\begin{center}
\begin{tabular}{|c|c|}
\hline \rule{0mm}{5mm}
 Isomultiplets & $B_{eq}$ \\[1mm]
\hline \rule{0mm}{7mm} $\displaystyle
\begin{array}{c} {}^{-1/3}S_0 \\ {}^{-2/3}V_0 \end{array}$ & 
$\displaystyle \frac{1+r}{2+r}$ \\[3mm]
\hline \rule{0mm}{7mm} $\displaystyle \begin{array}{c}
{}^{-4/3}\tilde{S}_0 \\ {}^{-5/3}\tilde{V}_0 \end{array}$ & 
1   \\[1mm]
\hline \rule{0mm}{9mm}
$\displaystyle \begin{array}{c} {}^{+2/3}S_1, \\ {}^{+1/3}V_1,
\end{array}$ 
$\displaystyle \begin{array}{c} {}^{-1/3}S_1, \\ {}^{-2/3}V_1,
\end{array}$ 
$\displaystyle \begin{array}{c} {}^{-4/3}S_1 \\ {}^{-5/3}V_1
\end{array}$ & 
0, $\displaystyle \frac{1}{2}$, 1 \\[4mm]
\hline \rule{0mm}{9mm}
$\displaystyle \begin{array}{c} {}^{-2/3}S_{1/2}, \\ {}^{-1/3}V_{1/2}, 
\end{array}$ 
$\displaystyle \begin{array}{c} {}^{-5/3}S_{1/2} \\ {}^{-4/3}V_{1/2} 
\end{array}$ &
$\displaystyle \frac{r}{1+r}$, 1 \\[4mm]
\hline \rule{0mm}{9mm}
$\displaystyle \begin{array}{c} {}^{+1/3}\tilde{S}_{1/2}, \\
{}^{+2/3}\tilde{V}_{1/2}, \end{array}$
$\displaystyle \begin{array}{c} {}^{-2/3}\tilde{S}_{1/2} \\
{}^{-1/3}\tilde{V}_{1/2}\end{array}$ &
0, 1  \\[4mm]
\hline
\end{tabular}
\caption{\it Branching ratios for charged lepton channels of generic
  leptoquarks ($r = \lambda_R^2/\lambda_L^2$).}
\label{tabbr} 
\end{center}
\end{table}

The formulae and prescriptions summarized in this section are
implemented in the Monte Carlo event generator {\tt LQPAIR}\footnote{The
  program was developed from an earlier generator {\tt LQ2} by D.\ 
  Gingrich \cite{gingrich}. The code is available from {\tt
    http://www.desy.de/\~\,hspiesb/lqpair.html}.}.  This program
simulates production and decay of leptoquarks and is interfaced to
parton shower and hadronization routines of {\tt JETSET}.  In the case
of initial state radiation, events containing a bremsstrahlung photon
are generated according to Eq.\ (\ref{dsigmac}).  The logarithm $\ln
\left(4E^2/m_e^2\right)$ in Eq.\ (\ref{ISRrad}) arises from the
  integration over transverse momenta of the emitted bremsstrahlung
  photon or, equivalently, over the emission angle $\theta_{\gamma}$ of
  the photon with respect to the incoming electron (positron) beam:
\begin{equation}
\ln \frac{4E^2}{m_e^2} = \int_{-1}^{+1} d\cos\theta_{\gamma}
\frac{E}{E-p\cos\theta_{\gamma}}
\label{thetadis}
\end{equation}
where $E \gg m_e$ is assumed. Therefore, transverse momenta are
generated according to the integrand in Eq.\ (\ref{thetadis}).

\section{Experimental Considerations}

\subsection{Detector Models}

In addition to beam- and bremsstrahlung, we shall take into account
detector effects such as limited acceptance and resolution.
Geometrical acceptance cuts (beam hole) reduce the cross section, in
particular in the forward and backward regions, which is important in
cases where large $t$-channel contributions (due to large Yukawa
couplings) are present.  Furthermore, measurement uncertainties
leading to a smearing of the lepton and hadron four-momenta
deteriorate the reconstruction of jets and thus make the
reconstruction of leptoquark masses less certain.

For our study we use two simple detector models interfaced to the
leptoquark event generator. The main components are a tracking device
and electromagnetic and hadronic calorimetry. Detector effects are
described by Gaussian smearing of particle four-momenta. The
single-particle detection efficiency is set to 98\,\% and the
threshold energy to $0.3$, $0.15$, and $0.6$ GeV for electromagnetic
particles, charged hadrons, and neutral hadrons, respectively. Table
\ref{tabsmear} lists the main detector parameters for the two models.
A more detailed discussion of the detector models will be given
elsewhere \cite{smear30}.  Our final results refer to the dedicated
1\,TeV detector. In selected cases we also compare the latter with an
LEP/SLC-type detector.

\begin{table}[htbp]
\begin{center}
\begin{tabular}{|c|l|l|}
\hline \rule{0mm}{6mm}
      & 1 TeV detector & LEP/SLC detector \\[1mm]
\hline
Electromagnetic calorimeter &                          & \\
$\displaystyle
\frac{\sigma(E)}{E} = \frac{A_{\rm em}}{\sqrt{E}} + B_{\rm em}$ &
                     $A_{\rm em} = 0.10$, $B_{\rm em} = 0.01$ & 
                     $A_{\rm em} = 0.20$, $B_{\rm em} = 0.01$ \\[1em] 
\hline
Hadronic calorimeter        &                          & \\
$\displaystyle
\frac{\sigma(E)}{E} = \frac{A_{\rm had}}{\sqrt{E}} + B_{\rm had}$ &
                     $A_{\rm had} = 0.50$, $B_{\rm had} = 0.04$ &  
                     $A_{\rm had} = 0.90$, $B_{\rm had} = 0.02$ \\[1em] 
\hline
Tracking (angular dependent)&                            &      \\
$\displaystyle
\frac{\sigma(p_{x,y})}{p_{x,y}^2} = P_R$         &
                     $P_R(90^{\circ}) = 2\cdot10^{-4}$   &     
                     $P_R(90^{\circ}) = 6\cdot10^{-4}$   \\[1em]
\hline
Multiple scattering  &                            &      \\
$\displaystyle
\frac{\sigma(p_{x,y,z})}{p_{x,y,z}^2} = 
 \frac{P_{MS}}{p_{x,y,z}\sqrt{\sin\theta}}$       &
                     $P_{MS} = 0.0015$            &
                     $P_{MS} = 0.0050$      \\[1em]
\hline \rule{0mm}{6mm}
Beam hole            &   $8.1^{\circ}$       &  $15^{\circ}$ \\[1mm]
\hline
\end{tabular}
\caption{\it Detector properties.}
\label{tabsmear}
\end{center}
\end{table}

\subsection{Search Strategies}

Leptoquarks decay into a lepton, either a charged one or a neutrino,
and a quark which is observed as a jet of hadrons. Therefore the final
states to be searched for are 
\begin{equation}
{\rm (I)}~~ \ell^+ \ell^- + 2 {\rm jets}, 
\label{signalI}
\end{equation}
\begin{equation}
{\rm (II)}~~ \ell^{\pm} + 2 {\rm jets} + p_{\rm miss},
\label{signalII}
\end{equation}
\begin{equation}
{\rm (III)}~~ 2 {\rm jets} + p_{\rm miss}
\label{signalIII}
\end{equation}
where the lepton and quark flavors are assumed to belong to one given
generation. We shall focus on the first generation, although most of
our results should also hold for second generation leptoquarks. 

Search I is straightforward. One has to identify two charged leptons
in the event. The events should contain enough hadronic energy in
order to allow the use of a jet algorithm to group the final state
hadrons into two well-defined jets. Next, one combines one of the
leptons with one of the jets. That combination which gives the
smallest difference in the invariant lepton-jet masses is accepted
and the lepton-jet masses are associated with the leptoquark mass.

Search II requires the identification of events with one charged lepton,
two jets, missing momentum and the determination of $p_{\rm miss}$.
Assuming the observed missing momentum to be carried by one single
neutrino, the event analysis can be performed in complete analogy to
search I .

In search III, the signal is characterized by missing momentum and
missing mass carried away by the unobserved neutrino pair. This is the
most difficult case. Obviously, the measurements which can be carried
out are not sufficient to reconstruct the leptoquark mass.

For clean event identification and precise mass reconstruction we
require the following cuts on energies and transverse momenta in
search I to III:
\begin{itemize}
\item[(I)] two charged leptons with transverse momentum $p_T^l \geq
  20$ GeV, missing transverse momentum $p_T^{miss} \leq 25$ GeV, two
  jets with energies $E_j \geq 10$ GeV, and total visible energy
  $E_{vis} \geq 0.9\sqrt{s}$;
\item[(II)] one, and only one, charged lepton with $p_T^l \geq 20$
  GeV, $p_T^{miss} \geq 25$ GeV, two jets with $E_j \geq 10$ GeV, and
  $E_{vis} \geq 0.6\sqrt{s}$;
\item[(III)] no charged lepton with $p_T^l \geq 20$ GeV, $p_T^{miss}
  \geq 25$ GeV, two jets with transverse momentum $p_T^j \geq 75$ GeV
  each, and total hadronic energy $E_{had} \leq 300$ GeV. The upper
  limit on $E_{had}$ corresponds to a lower limit for the missing mass
  of the neutrino pair.
\end{itemize}
These general requirements are supplemented by additional cuts to
suppress the background processes which are discussed in the next
subsection.

\subsection{Background Processes}

Vector boson pair-production is an obvious source of final states of the
kind Eqs.\ (\ref{signalI}) to (\ref{signalIII}) within the standard
model:
\begin{equation}
e^+e^- \rightarrow ZZ, 
\end{equation}
\begin{equation}
e^+e^- \rightarrow W^+W^-
\end{equation}
with one $Z$ ($W$) decaying into a lepton pair and the other one into
two jets.  More generally, one has to consider four-fermion production
\begin{equation}
e^+e^- \rightarrow l_1 \bar{l_2} q_3 \bar{q_4}
\end{equation}
comprising also single-resonant processes like $e^+e^- \rightarrow
Z\nu\nu, Z\ell\ell, W\ell\nu$ with $Z \rightarrow q\bar{q}$ and $W
\rightarrow q_1\bar{q}_2$, as well as completely non-resonant
processes.  Whereas on-shell $W$ and $Z$ production can be suppressed
efficiently by rejecting events with pairs of final state leptons or
jets having a mass close to $M_W$ or $M_Z$, suppression of off-shell
production requires additional cuts. We implement them as cuts on the
various invariant masses $M_{l_i j_k}$. 

Apart from the irreducible background due to final states containing the
same particles as the signal events, there is in addition some reducible
background due to processes with the same visible particles produced by
different final states. In particular, $b$-quarks and $\tau$-leptons
have some probability to produce final states with an energetic
electron, that may mimic the decay products of a leptoquark.

A second important source of background is $t\bar{t}$ production:
\begin{equation}
e^+e^- \rightarrow t\bar{t}
\end{equation}
with the top-quarks decaying into $b$ and $W$ and the latter decaying
leptonically.

The cuts devised to suppress these background processes are summarized
below.  The indices $j_1$ and $j_2$ label the two jets ordered by their
energies, i.e., $E_{j_1} > E_{j_2}$. $l_1$ and $l_2$ denote the leptons
ordered in such a way that
\begin{equation}
\left|M_{l_1j_1} - M_{l_2j_2}\right| < \left|M_{l_1j_2} -
  M_{l_2j_1}\right|. 
\end{equation}
With this convention, the pairs $l_1j_1$ and $l_2j_2$ are candidates for
the decay products of a leptoquark. In search (I) to (III) we
require:
\begin{itemize}
\item[(I)] $|M_{l_1l_2} - M_Z| \geq 10$ GeV and $|M_{j_1j_2} - M_{W,Z}|
  \geq 10$ GeV; \\ $M_{l_1l_2} \geq 20$ GeV, $M_{l_1j_2} \geq 20$ GeV,
  $M_{l_2j_1} \geq 20$ GeV.
\item[(II)] $|M_{l_1l_2} - M_W| \geq 20$ GeV and $|M_{j_1j_2} -
  M_{W,Z}| \geq 20$ GeV; \\ $M_{l_1l_2} \geq 20$ GeV, $M_{l_1j_2} \geq
  20$ GeV, $M_{l_2j_1} \geq 20$ GeV;\\ $E_{had} \geq 150$ GeV.
\item[(III)] $|M_{l_1l_2} - M_Z| \geq 20$ GeV and $|M_{j_1j_2} -
  M_{W,Z}| \geq 20$ GeV; \\ $M_{j_1j_2} \leq 400$ GeV.
\end{itemize}

\begin{table}[htb]
\begin{center}
\begin{tabular}{|l|c|c|c|c|c|c|}
\hline \rule{0mm}{5mm}
      & \multicolumn{3}{|c|}{$\sqrt{s}=500$ GeV}
      & \multicolumn{3}{|c|}{$\sqrt{s}=800$ GeV} \\[1mm] 
\rule{0mm}{5mm}
      & \multicolumn{3}{|c|}{${\cal L} = 20$ fb$^{-1}$}
      & \multicolumn{3}{|c|}{${\cal L} = 50$ fb$^{-1}$} \\[1mm]
\hline \rule{0mm}{5mm}
      ~source
      & I & II & III & I & II & III \\[1mm] 
\hline
\hline \rule{0mm}{13mm}
$\displaystyle\begin{array}{l}
e^+e^-q\bar{q},~q=u,d,s,c,b \\
e^-\nu_e\bar{u}d + {\rm c.c.} \\
e^-\nu_e\bar{c}s + {\rm c.c.}
\end{array}$
            & 11.9 & 14.8 & 266  & 15.0 &  39.3 & 1550 \\[9mm]
\hline \rule{0mm}{8mm}
$\displaystyle\begin{array}{l}
b\bar{b}f\bar{f},~f\neq e
\end{array}$
            &  --  &  1.4 & 495  &  --  &   1.3 & 2410 \\[4mm]
\hline \rule{0mm}{11mm}
$\displaystyle\begin{array}{l}
\tau^+\tau^-f\bar{f},~f\neq b \\
\tau^{\pm}\stackrel{(-)}{\nu_{\tau}}q\bar{q'} 
\end{array}$
            &  0.4 &  1.3 & 888  &  0.6 &   5.1 &  320 \\[7mm]
\hline \rule{0mm}{5mm}
~other 4f    &  --  &  0.1 & 371  &  --  &   8.0 & 1250 \\[1mm]
\hline
\hline \rule{0mm}{5mm}
~sum of 4f   & 12.3 & 17.6 & 2010 & 15.6 &  53.7 & 5540 \\[1mm]
\hline
\hline \rule{0mm}{5mm}
~$t\bar{t}$  &  0.7 & 132  & 527  &  1.5 &  91.6 &   97.8 \\[1mm]
\hline
\hline \rule{0mm}{5mm}
~total       & 13.0 & 150 & 2537  & 17.1 & 145.3 & 5630 \\[1mm]
\hline
\end{tabular}
\caption{\it Number of background events surviving the cuts in the 
  leptoquark searches (I) to (III) at $\protect\sqrt{s} = 500$ GeV
  (${\cal L} = 20$ fb$^{-1}$) and 800 GeV (${\cal L} = 50$
  fb$^{-1}$).}
\label{tabbg}
\end{center}
\end{table}

Table \ref{tabbg} indicates the number of remaining events from
various reducible and irreducible background processes. The estimates
for four-fermion final states have been obtained with the help of {\tt
  WPHACT} \cite{wphact}, those for $t\bar{t}$ production with {\tt
  PYTHIA} \cite{pythia}. Beamstrahlung and ISR are not included here.
These corrections should be negligible for the present purpose.

In search I, the dominant background is due to on- and off-shell boson
pair-production with $e^{\pm}$ in the final state, i.e., $e^+e^-
\rightarrow e^+e^- q\bar{q}$ and $e^+e^- \rightarrow e^{\pm}\nu
q\bar{q'}$.  The rate is much higher than at LEP2 energies estimated
in a previous study of leptoquark production \cite{lep2}. Whereas
$t\bar{t}$ production is not a problem for search I, it makes the
search for leptoquarks in channel II very difficult. In search III,
the number of background events surviving the cuts is very large,
since the final state cannot be reconstructed completely. In
particular, final states with $b$-quarks and $\tau$'s contribute a
large fraction. It seems unlikely that in this channel one can expect
more than a consistency check of the searches I and II.

Due to their huge cross sections, two-photon processes have also to be
considered as a potential source of background although four-fermion
final states only emerge at higher orders. We checked for a sample of
$2\cdot 10^5$ $\gamma\gamma$ events generated with the help of the
event generator {\tt PYTHIA} (version 5.722) that no event passed the
cuts for searches I and II, whereas in search III a negligibly small
number of events remained.

\section{Results}

In the following, we restrict ourselves to generic leptoquarks of the
first generation which decay into $e^{\pm}$ or $\nu_e$ and a
jet. Except where stated differently we assume
\begin{equation} 
\lambda_L \ll 1, ~~~~ \lambda_R \ll 1,
\end{equation}
so that the Yukawa couplings have negligible effects on the production
cross sections. Note that in this case the production rates are
actually independent of the generation quantum number.  However, the
decay products of third generation leptoquarks contain the heavy
flavors $\tau$, $b$ and $t$ which in turn decay into lighter leptons
and jets.  Hence, the search strategies explained in section 3 are not
appropriate.

\subsection{Signal Distributions}

\begin{figure}[htbp] 
\unitlength 1mm
\begin{picture}(160,205)
\put(2,120){
\epsfxsize=8cm
\epsfysize=10.5cm
\epsfbox{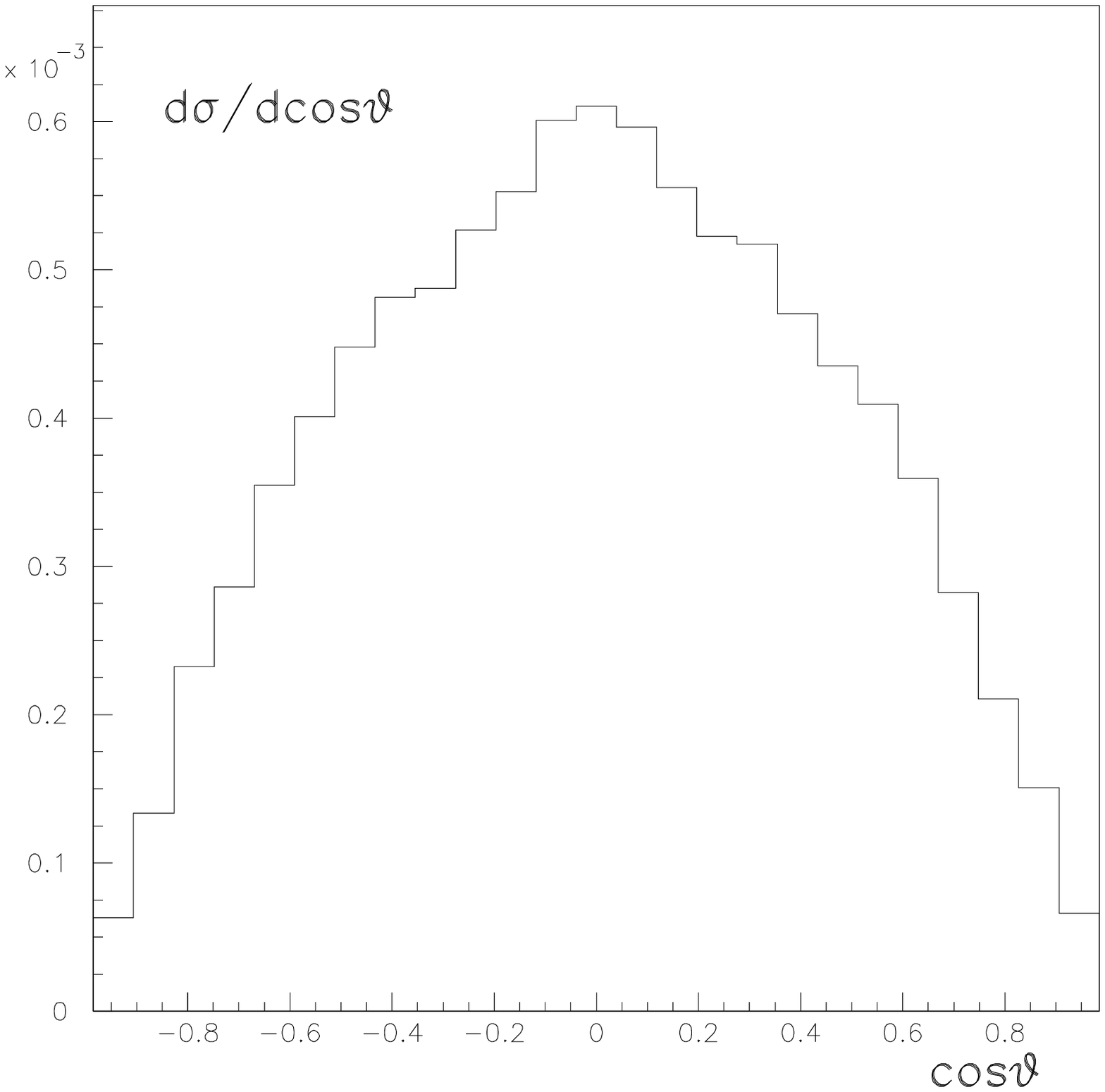}
}
\put(41,138){{\rm (a)}}
\put(82,120){
\epsfxsize=8cm
\epsfysize=10.5cm
\epsfbox{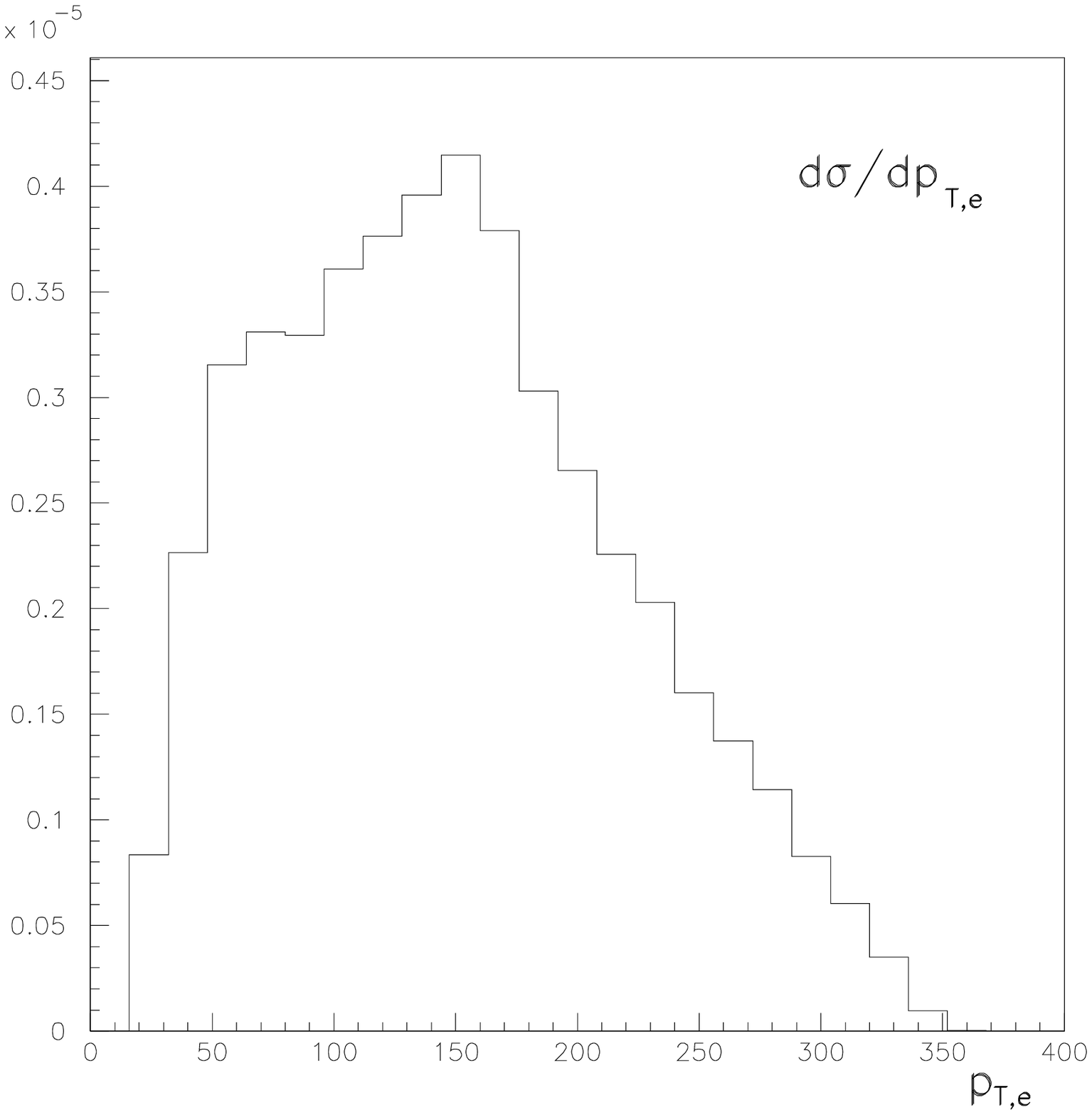}
}
\put(120,138){{\rm (b)}}
\put(2,51){
\epsfxsize=8cm
\epsfysize=10.5cm
\epsfbox{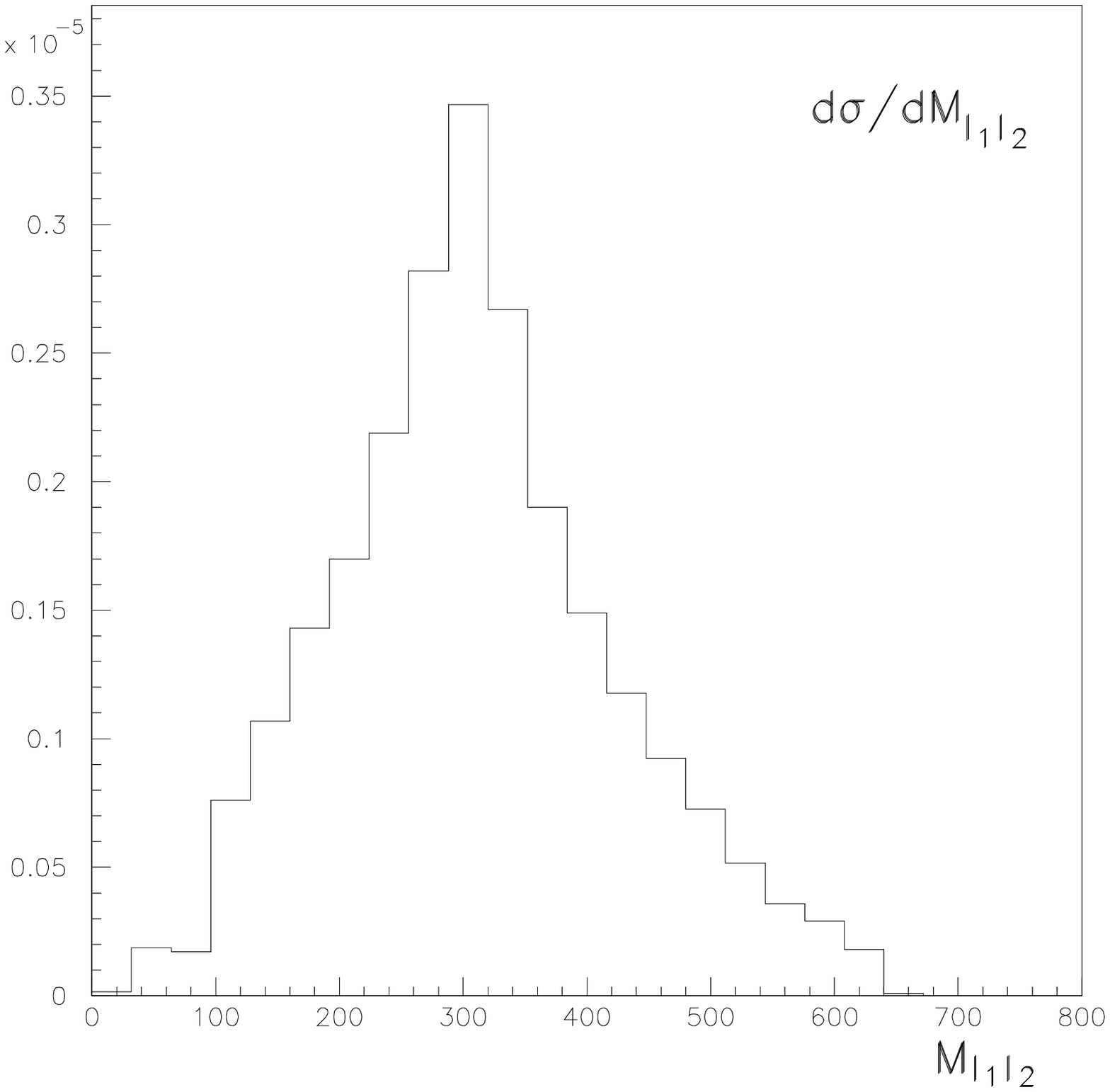}
}
\put(41,69){{\rm (c)}}
\put(82,51){
\epsfxsize=8cm
\epsfysize=10.5cm
\epsfbox{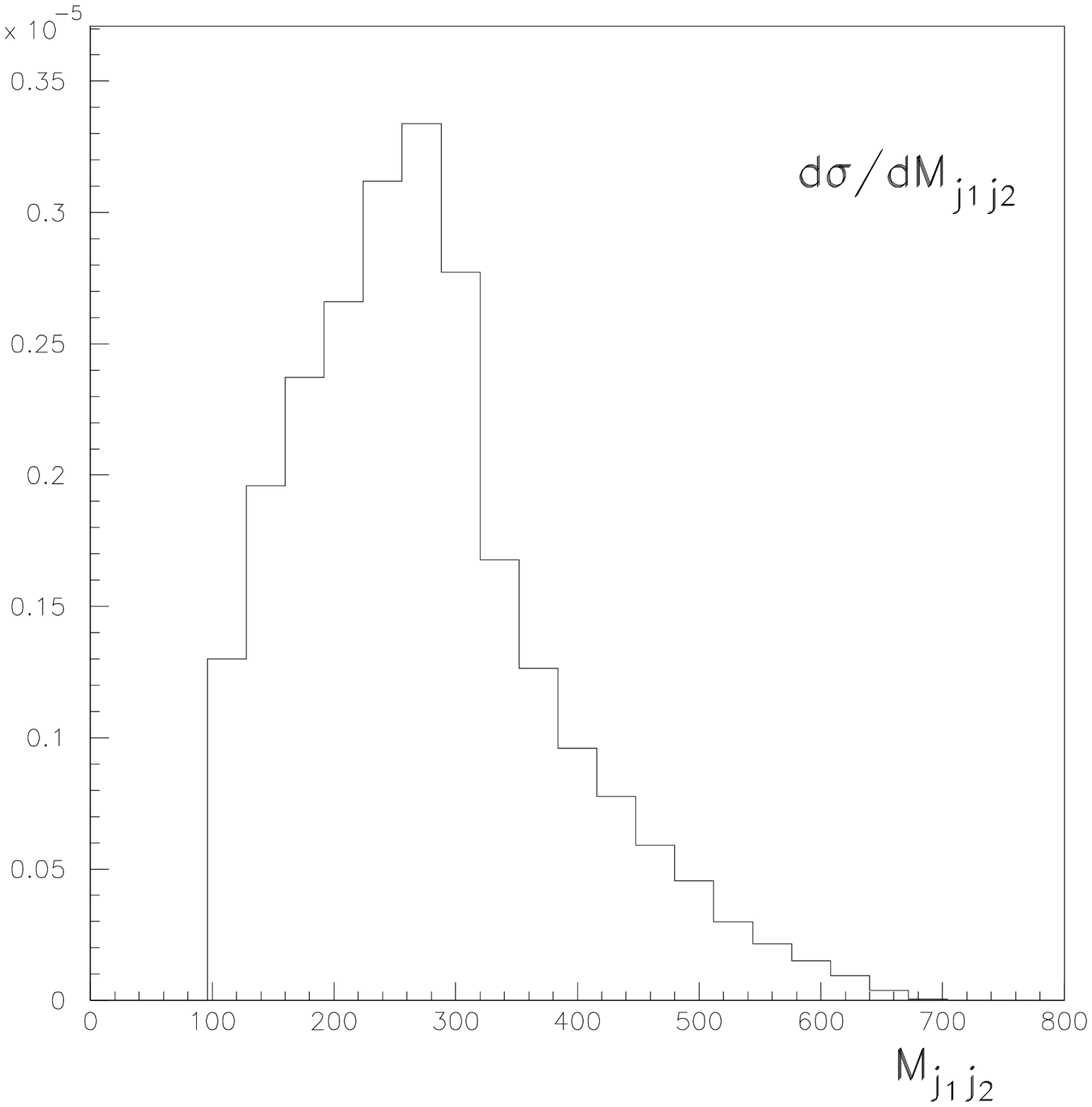}
}
\put(120,69){{\rm (d)}}
\put(2,-18){
\epsfxsize=8cm
\epsfysize=10.5cm
\epsfbox{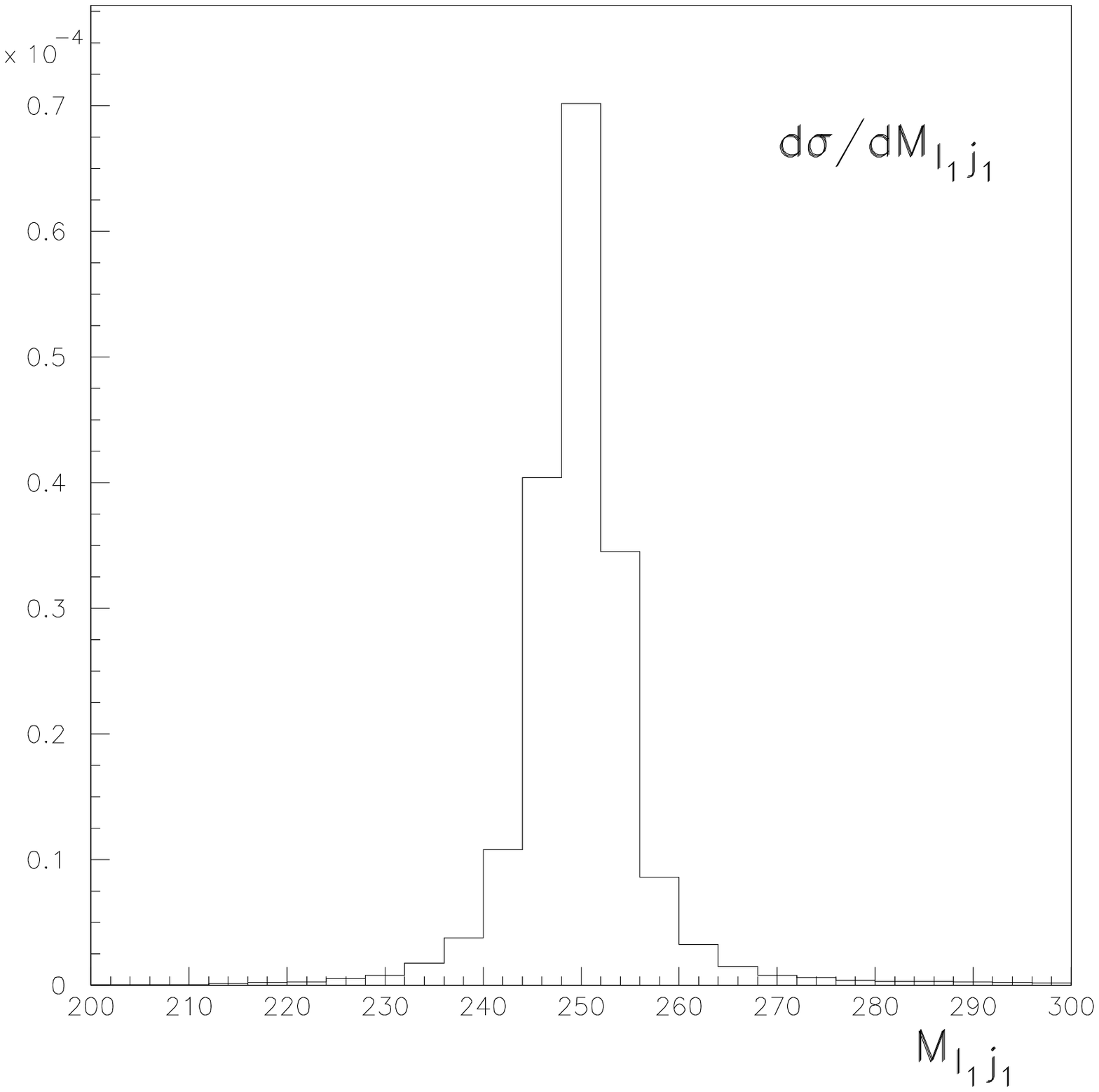}
}
\put(41,0){{\rm (e)}}
\put(82,-18){
\epsfxsize=8cm
\epsfysize=10.5cm
\epsfbox{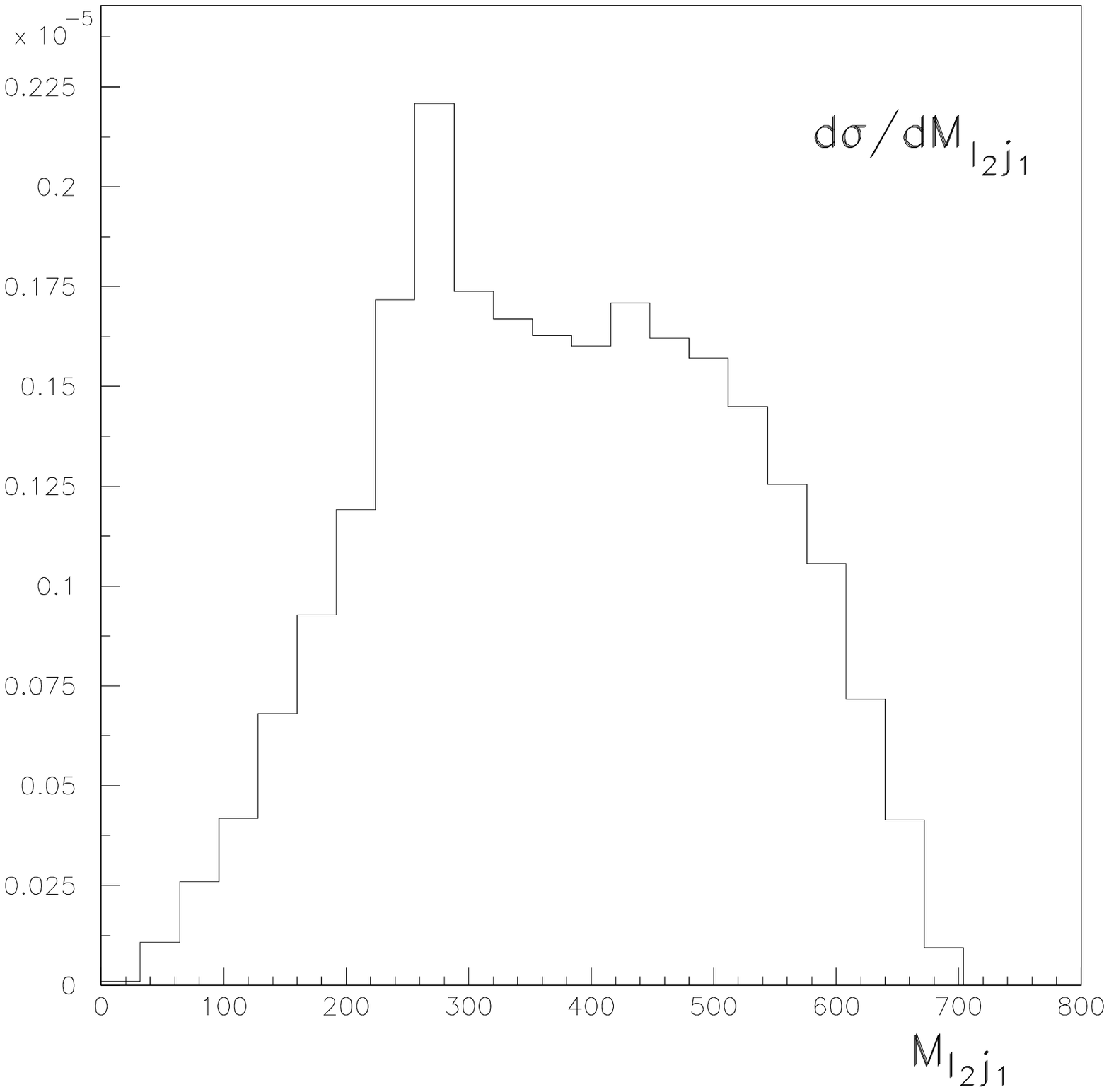}
}
\put(120,0){{\rm (f)}}
\end{picture}
\caption{\it Event distributions for the process $e^+e^- \rightarrow
  {}^{-1/3}S_1 + {}^{1/3}\bar{S}_1 \rightarrow e^{\pm}
  \stackrel{(-)}{\nu_e} jj$ for $M = 250$ GeV at $\protect\sqrt{s} =
  800$ GeV (search II).  Masses are given in GeV and cross sections in
  pb. The definition of kinematic variables is explained in subsection
  3.3.}
\label{figkindis}
\end{figure}

Figure \ref{figkindis} illustrates some characteristic distributions
generated from the class II process $e^+e^- \rightarrow {}^{-1/3}S_1 +
{}^{1/3}\bar{S}_1 \rightarrow e^{\pm} \stackrel{(-)}{\nu_e} jj$ for
$M=250$ GeV and $\sqrt{s} = 800$ GeV in a high statistics Monte Carlo
run.  Beamstrahlung, ISR, detector effects and cuts are taken into
account as described above. In Fig.\ \ref{figkindis}a one sees
the $\sin^2 \theta$-distribution characteristic for the production of
scalar particles which can nicely be reconstructed under the given
assumptions.  Figure \ref{figkindis}b displays the distribution in the
transverse momentum of the decay electron (positron), which is
practically the same for the missing transverse momentum and the jet
transverse momentum.  The invariant-mass distributions of the lepton
and the jet pairs shown in Figs.\ \ref{figkindis}c and
\ref{figkindis}d have their maximum at about 300 GeV, while the
leptoquark signal is clearly visible in the distribution of $M_{l_1
  j_1}$ of Fig.\ \ref{figkindis}e.  The tails in this distribution
extending to smaller and higher masses are small.  From Fig.\ 
\ref{figkindis}f we see that the identification of the correct
combination of leptons and jets is not perfect: the mass distribution
of the `wrong' combination shows some excess of events in the region
of $M_{l_2j_1} \simeq 250$ GeV $=M$.

\subsection{Detection Efficiencies}

The cuts described in subsections 3.2 and 3.3 are considered a
reasonable compromise between the two conflicting requirements of
background suppression and preservation of the signal rate. It is
certainly possible to further optimize them. Particularly worthwhile
may be some fine-tuning for different ranges of center-of-mass
energies and leptoquark masses.

\begin{table}[htb]
\begin{center}
\begin{tabular}{|r|c|c|c|c|}
\hline \rule{0mm}{5mm}
I ~~~~~~~~ & number of events       & signal efficiency (\%)
           & standard               & mass shift              \\
$M=225$ GeV
      &   $N_{\rm events}/B_I$      &           
      & deviation                   & (GeV)                   \\
$\sqrt{s}=500$ GeV
      & (${\cal L} = 20$ fb$^{-1}$) &                         
      & (GeV)                       &                         \\[1mm]
\hline
\hline \rule{0mm}{5mm}
${}^{-1/3}S_0$ & 9 / 8 & 26.5 / 23.8 & 4.8 / 5.9 & $-3.0$ / $-3.1$ \\
\hline
${}^{-4/3}S_1$ & 259 / 220 & 27.2 / 23.1 & 5.0 / 5.9 & $-2.9$ / $-3.3$ \\
\hline
${}^{-2/3}V_0$ & 313 / 269 & 27.2 / 23.4 & 5.0 / 5.9 & $-3.0$ / $-3.2$ \\
\hline
${}^{-5/3}V_1$ & 2640 / 2300 & 28.1 / 24.5 & 4.5 / 5.8 & $-2.9$ / $-3.2$ 
\\[1mm]
\hline
\hline
I ~~~~~~~~ &                      & 
      &                      & \\
$M=350$ GeV
      & $N_{\rm events}/B_I$  &                      &               
      &                       \\
$\sqrt{s}=800$ GeV
      & (${\cal L} = 50$ fb$^{-1}$) &                      &               
      &                      \\
\hline \rule{0mm}{5mm}
${}^{-1/3}S_0$ & 19 / 15 & 36.8 / 28.7 & 5.1 / 6.6 & $-3.3$ / $-4.2$ \\
\hline
${}^{-4/3}S_1$ & 492 / 378 & 35.9 / 27.8 & 5.2 / 6.8 & $-3.2$ / $-4.2$ \\
\hline
${}^{-2/3}V_0$ & 608 / 486 & 35.1 / 28.1 & 5.2 / 6.6 & $-3.2$ / $-4.4$ \\
\hline
${}^{-5/3}V_1$ & 5080 / 3780 & 36.3 / 27.0 & 5.1 / 6.6 & $-3.3$ / $-4.1$ 
\\[1mm]
\hline
\hline
II ~~~~~~~~ &                      & 
      &                      & \\
$M=225$ GeV
      & $N_{\rm events}/B_{II}$ &                      &              
      &                      \\
$\sqrt{s}=500$ GeV
      & (${\cal L} = 20$ fb$^{-1}$) &                      &              
      &                      \\
\hline \rule{0mm}{5mm}
${}^{-1/3}S_0$ & 8 / 7 & 24.0 / 19.5 & 6.5 / 8.0 & $+1.3$ / $+1.6$ \\
\hline
${}^{-2/3}V_0$ & 269 / 234 & 23.4 / 20.4 & 6.7 / 7.3 & $+1.6$ / $+0.9$ 
\\[1mm]
\hline
\hline
II ~~~~~~~~ &                      & 
      &                      & \\
$M=350$ GeV
      & $N_{\rm events}/B_{II}$ &                      & 
      &                      \\
$\sqrt{s}=800$ GeV
      & (${\cal L} = 50$ fb$^{-1}$) &                      & 
      &                      \\
\hline \rule{0mm}{5mm}
${}^{-1/3}S_0$ & 16 / 12 & 29.7 / 23.3 & 5.9 / 6.7 & $-0.8$ / $-1.3$ \\
\hline
${}^{-2/3}V_0$ & 488 / 397 & 28.2 / 23.0 & 5.8 / 6.9 & $-0.7$ / $-1.5$ 
\\[1mm]
\hline
\hline
\end{tabular}
\caption{\it Event number, signal efficiency, difference in  
  reconstructed and nominal leptoquark mass, and width of the
  reconstructed mass peak for two scenarios and in two search
  channels.  The expected numbers of events in column 2 are divided by
  the respective branching ratios $B_I = B_{eq}^2$, $B_{II} = 2 B_{eq}
  B_{\nu q}$. The first number given in each column refers to a
  dedicated 1 TeV detector, the second one to a LEP/SLC-type
  detector.}
\label{tabeffs}
\end{center}
\end{table}

Table \ref{tabeffs} gives an overview of the potential in the leptoquark
searches I and II. Column 2 displays the number of events observed
divided by the respective branching ratios $B_{I} = B_{eq}^2$ and
$B_{II} = 2 B_{eq} B_{\nu q}$ or, equivalently, the number of
leptoquarks produced and surviving the cuts in channel I and II,
respectively.  One can see that in both cases the signal efficiency is
reasonably large, between 20\,\% and 40\,\%.  There is no significant
difference in the efficiencies for scalar and vector leptoquarks.
Furthermore, we find that for a LEP/SLD-type detector the reconstruction
efficiency is lower by typically 20\,\% relative to the dedicated 1 TeV
detector, depending on the channel and the leptoquark mass. This
reduction originates mainly from the different size of the beam-holes
(see Table 4).

The results in Table \ref{tabeffs} are obtained for negligibly small
Yukawa couplings, i.e., for leptoquarks with very small natural widths
(see Eq.\ (\ref{widths})).  Beamstrahlung, initial state radiation, and
in particular hadronization effects lead to a considerable broadening of
the observed leptoquark mass distribution.  In order to estimate the
mass resolution, we fitted the reconstructed mass distribution to a
Gaussian. Column 4 of the table shows the standard deviation of such a
fit.  Note that the precision with which the peak position of the
reconstructed mass distribution can be determined is, in most cases,
better by a factor of 5 to 10, i.e., $\delta M = 0.5$ to $0.8$ GeV.
From column 4 and Eq.\ (\ref{widths}) one clearly sees that a
measurement of the mass distribution of leptoquark decay products will
not allow to determine the intrinsic leptoquark width for accessible
masses and values of $\lambda_{L,R}$ allowed by the existing bounds.

Finally, Table \ref{tabeffs} reveals a systematic shift of the
reconstructed leptoquark mass. Again, this is dominantly due to
hadronization effects and depends strongly on the algorithm used for
jet reconstruction (see for example \cite{jetalgo}). We compared the
JADE and Durham schemes with E, E0, P, and P0 recombination and found
that the Durham-P scheme leads to the smallest mass shifts. Other jet
algorithms induce considerably larger shifts, for example, the JADE-E
scheme a shift up to 10 GeV.  Therefore, the Durham-P algorithm was
used in the subsequent analysis.

In channel III, no mass reconstruction is possible. The signal
efficiency ranges from 40 to 45$\,\%$ at $\sqrt{s} = 500$ GeV and from
7 to 15$\,\%$ at $\sqrt{s} = 800$ GeV.

\subsection{Sensitivity Limits}

From the above it becomes clear that the discovery potential in the
three channels is very different. In other words, for a given
leptoquark species, the observability strongly depends on the
branching ratio for the decay into a charged lepton, here $e^{\pm}$.
As can be seen from Table \ref{tabbr}, this branching ratio is fixed
for all states with the exception of ${}^{-1/3}S_0$, ${}^{-2/3}V_0$,
${}^{-2/3}S_{1/2}$, and ${}^{-1/3}V_{1/2}$. In the latter cases, the
branching ratio is determined by the ratio $r = \lambda_R^2 /
\lambda_L^2$. Below we shall present results for $r = 0, 1$ and
$\infty$, corresponding to $B_{eq} = 1/2$, $2/3$, and 1.

A first estimate of the sensitivity limits can be based on the total
number of events. Requiring a $5\sigma$ effect, we determine the values
of $M$ for which the number of signal events is equal to or larger than
$5\sqrt{N_{bg}}$ where $N_{bg}$ is the total number of background
events. This is a sensible discovery criterion for search I where the
number of background events is small. The results are collected in Table
\ref{tablimits} for all leptoquark species of Table \ref{tabprop}.  In
search I, scalar leptoquarks can be discovered for masses between 80 and
97\,\% of $\sqrt{s}/2$ except for the states ${}^{-1/3}S_0$ and
${}^{-1/3}S_1$ where only $70\,\%$ of $\sqrt{s}/2$ can be reached for
purely left-handed couplings. For vector leptoquarks the mass reach is
always larger than 95\,\% of $\sqrt{s}/2$.  In searches II and III, the
discovery limits are worse. In these channels, some scalars cannot be
detected at all for masses above 100 GeV.  It should also be noted that
small, but non-negligible Yukawa couplings \cite{BluRu} and anomalous
couplings \cite{bbk} can lower the production cross sections and
discovery limits in comparison to Table \ref{tablimits}.

\begin{figure}[htbp] 
\unitlength 1mm
\begin{picture}(160,80)
\put(-8,-13){
\epsfxsize=9cm
\epsfysize=12cm
\epsfbox{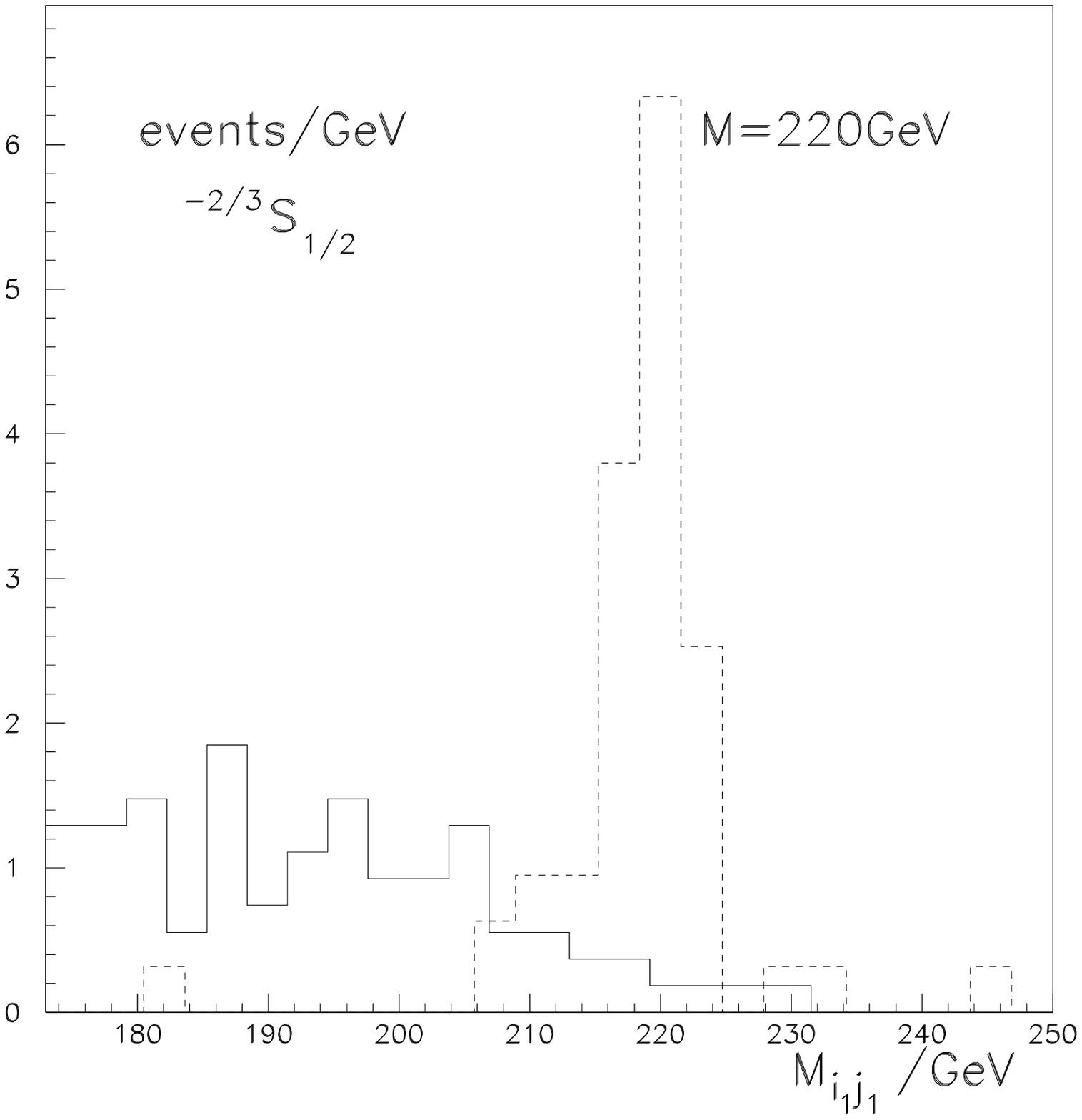}
}
\put(36,5){{\rm (a)}}
\put(77,-13){
\epsfxsize=9cm
\epsfysize=12cm
\epsfbox{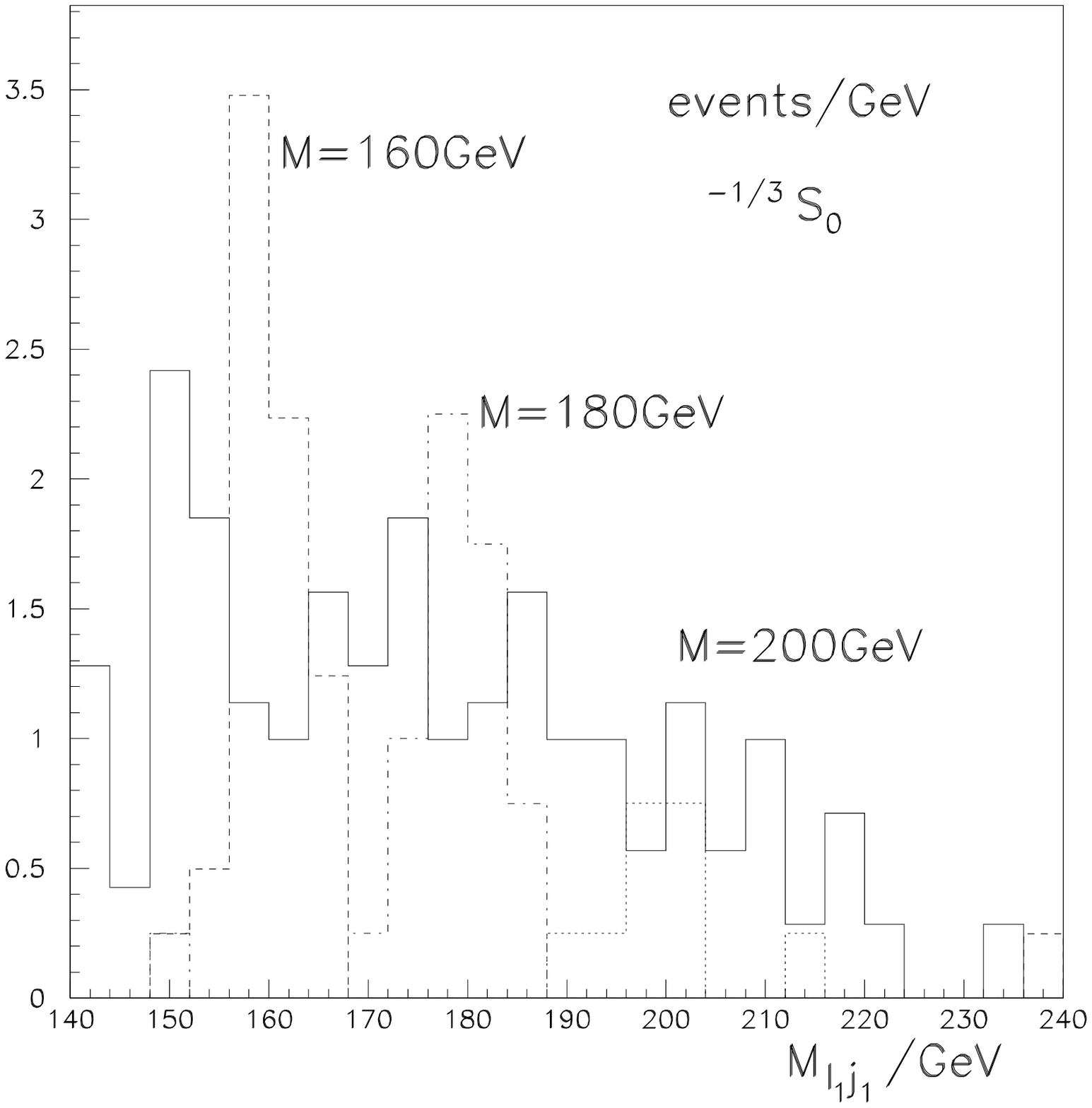}
}
\put(124,5){{\rm (b)}}
\end{picture}
\caption{\it Signal and background distributions in the invariant
  mass $M_{l_1j_1}$ for the channel $e^{\pm}\nu + 2$jets and
  $\protect\sqrt{s} = 500$ GeV, ${\cal L} = 20$ fb${}^{-1}$: (a)
  ${}^{-2/3}S_{1/2}$ production, (b) ${}^{-1/3}S_0$ production and
  $M=160$ GeV (dashed), $M=180$ GeV (dash-dotted) and $M=200$ GeV
  (dotted).  The full histograms show the dominant background from
  $t\bar{t}$ production.}
\label{figsigback}
\end{figure}

A more refined analysis is possible if the number of events is
sufficiently large to investigate the mass distributions. The signal is
expected to stick out as a prominent peak over a flat background
distribution. From an investigation of the mass distribution one would
obtain a higher significance for leptoquarks with the masses shown in
Table \ref{tablimits} than the global 5$\sigma$ assumed there.  This is
demonstrated in Fig.\ \ref{figsigback}a for a favorable case, while
Fig.\ \ref{figsigback}b illustrates a more difficult situation.  In the
latter, the 9 signal events expected for $M = 200$ GeV disappear in the
$t\bar{t}$ background, whereas the 33 signal events for $M = 160$ GeV
might still allow to identify an enhancement in the mass distribution.
This, however, requires a mass resolution better than 5 GeV.

\begin{table}[hbtp]
\begin{center}
\begin{tabular}{|l|l|c|c|c|c|c|c|}
\hline  \multicolumn{2}{|l|}{\rule{0mm}{5mm}}
      & \multicolumn{3}{|c|}{$\sqrt{s}=500$ GeV}
      & \multicolumn{3}{|c|}{$\sqrt{s}=800$ GeV} \\[1mm]
\hline  \multicolumn{2}{|r|}{\rule{0mm}{4.5mm} Search}      
      &   I &    II  &  III   &   I &    II  &   III \\
\hline
\multicolumn{2}{|r|}{\rule{0mm}{5mm} $5\sqrt{N_{bg}}$}
      &  18 &    61  &  251   &  21 &    60  &   375 \\[1mm]
\hline
States & $B_{eq}$ &
\multicolumn{6}{|c|}{\rule{0mm}{5mm} Mass reach in GeV} \\[1mm]
\hline
\hline \rule{0mm}{5mm}
${}^{-1/3}S_0$ & $2/3$ 
      &\hspace{1mm}202\hspace{1mm}
            &   $*$  &   $*$ &\hspace{1mm}318\hspace{1mm}
                                    &   $*$  & $*$ 
\\
               & $1/2$ 
      & 183 &   $*$  &   $*$  & 289 &   $*$  & $*$ 
\\
              & $1$ 
      & 217 &   --   &    --  & 350 &    --  &  -- 
\\[1mm]
\hline \rule{0mm}{5mm}
${}^{-4/3}\tilde{S}_0$ & $1$ 
      & 242 &   --   &    --  & 387 &    --  &  -- 
\\[1mm]
\hline \rule{0mm}{5mm}
${}^{2/3}S_1$ & $0$ 
      & -- &    --   &   225  &  -- &    --  &  275
\\[1mm]
\hline \rule{0mm}{5mm}
${}^{-1/3}S_1$ & $1/2$ 
      & 183 &   $*$  &   $*$  & 289 &    $*$ & $*$ 
\\[1mm]
\hline \rule{0mm}{5mm}
${}^{-4/3}S_1$ & $1$ 
      & 244 &   --   &    --  & 389 &    --  &  --
\\[1mm]
\hline \rule{0mm}{5mm}
${}^{-2/3}S_{1/2}$ & $1/2$ 
      & 230 &  221   &   179  & 369 &   359  &  $*$
\\
                   & $0$ 
      & --  &   --   &   218  &  -- &    --  &  239
\\
                   & $1$ 
      & 240 &   --   &   --   & 384 &    --  &  --
\\[1mm]
\hline \rule{0mm}{5mm}
${}^{-5/3}S_{1/2}$ & $1$
      & 244 &  --    &   --   & 389 &   --   &  --  
\\[1mm]
\hline \rule{0mm}{5mm}
${}^{1/3}\tilde{S}_{1/2}$ & $0$
      &  -- &  --    &  198   & --  &   --   & 146 
\\[1mm]
\hline \rule{0mm}{5mm}
${}^{-2/3}\tilde{S}_{1/2}$ & $1$
      & 237 &  --    &   --   & 379 &   --   &  --  
\\[1mm]
\hline
\hline \rule{0mm}{5mm}
${}^{-1/3}V_{1/2}$ & $1/2$ 
      & 241 &  237   &  220   & 385 &   380  & 266 
\\
                   & $0$ 
      &  -- &   --   &  236   &  -- &    --  & 326
\\
                   & $1$ 
      & 245 &   --   &   --   & 392 &    --  &  -- 
\\[1mm]
\hline \rule{0mm}{5mm}
${}^{-4/3}V_{1/2}$ & $1$
      & 247 &   --   &   --   & 395 &    --  &  -- 
\\[1mm]
\hline \rule{0mm}{5mm}
${}^{2/3}\tilde{V}_{1/2}$ & $0$
      & --  &   --   &  236   & --  &    --  & 326
\\[1mm]
\hline \rule{0mm}{5mm}
${}^{-1/3}\tilde{V}_{1/2}$ & $1$
      & 244 &   --   &   --   & 390 &    --  &  -- 
\\[1mm]
\hline \rule{0mm}{5mm}
${}^{-2/3}V_0$ & $2/3$
      & 241 &  233   &  195   & 385 &   373  & 200
\\
               & $1/2$
      & 238 &  234   &  212   & 380 &   376  & 244
\\
               & $1$
      & 244 &   --   &   --   & 390 &    --  &  -- 
\\[1mm]
\hline \rule{0mm}{5mm}
${}^{-5/3}\tilde{V}_0$ & $1$
      & 247 &   --   &   --   & 396 &    --  &  --  
\\[1mm]
\hline \rule{0mm}{5mm}
${}^{1/3}V_1$ & $0$
      & --  &   --   &  241   & --  &    --  & 352
\\[1mm]
\hline \rule{0mm}{5mm}
${}^{-2/3}V_1$ & $1/2$
      & 238 &  234   &  212   & 380 &   375  & 244
\\[1mm]
\hline \rule{0mm}{5mm}
${}^{-5/3}V_1$ & $1$
      & 248 &   --   &   --   & 396 &    --  &  --  
\\[1mm]
\hline
\end{tabular} 
\caption{\it Discovery limits for leptoquarks (masses in GeV) at 
  $\protect\sqrt{s} = 500$ GeV (${\cal L} = 20$ fb$^{-1}$) and
  $\protect\sqrt{s} = 800$ GeV (${\cal L} = 50$ fb$^{-1}$) requiring a
  5$\sigma$ effect. Dashes indicate cases where the corresponding search
  is not possible, $*$ means no sensitivity to masses above 100 GeV
  with the cuts considered.}
\label{tablimits}
\end{center}
\end{table}

Note that for Fig.\ \ref{figsigback} we have assumed $\lambda_L =
\lambda_R \ll 1$.  Vanishing $\lambda_L$ forbids class II final
states. For ${}^{-2/3}S_{1/2}$ (Fig.\ \ref{figsigback}a) this is also
the case for $\lambda_R = 0$, while for ${}^{-1/2}S_0$ the signal
distributions shown in Fig.\ \ref{figsigback}b have to be rescaled by
the factor $8/9$.  For small, but non-negligible Yukawa couplings, the
production rates may be either enhanced or reduced relative to the
case of negligible $\lambda_{L,R}$, whereas for sufficiently large
couplings the $t$-channel contribution always leads to larger cross
sections\footnote{The influence of anomalous couplings is discussed in
  Ref.\ \cite{bbk}.}.

\subsection{Leptoquark Couplings}

\begin{figure}[htb] 
\unitlength 1mm
\begin{picture}(160,85)
\put(-6,-20){
\epsfxsize=9cm
\epsfysize=13cm
\epsfbox{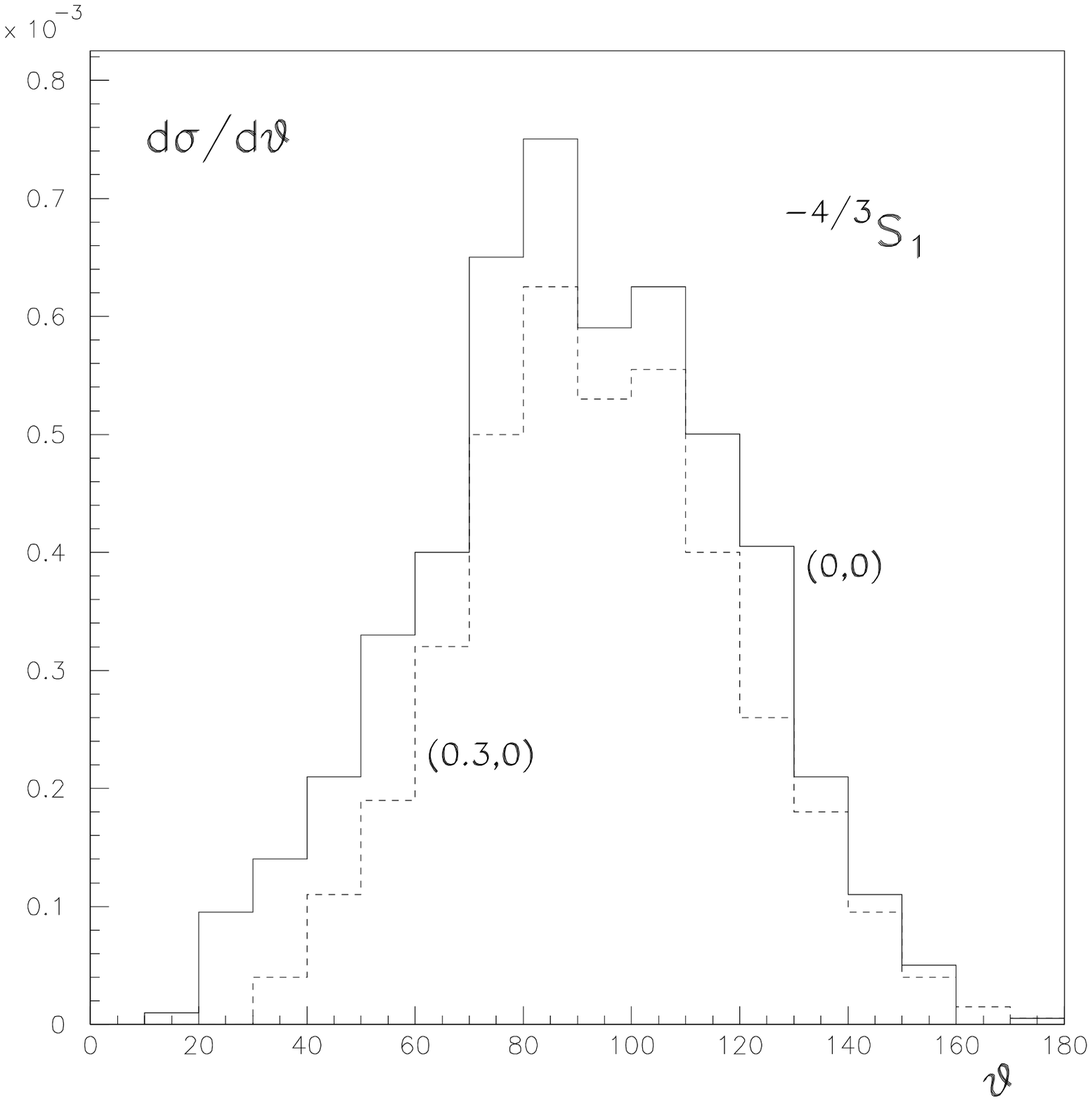}
}
\put(38,0){{\rm (a)}}
\put(78,-20){
\epsfxsize=9cm
\epsfysize=13cm
\epsfbox{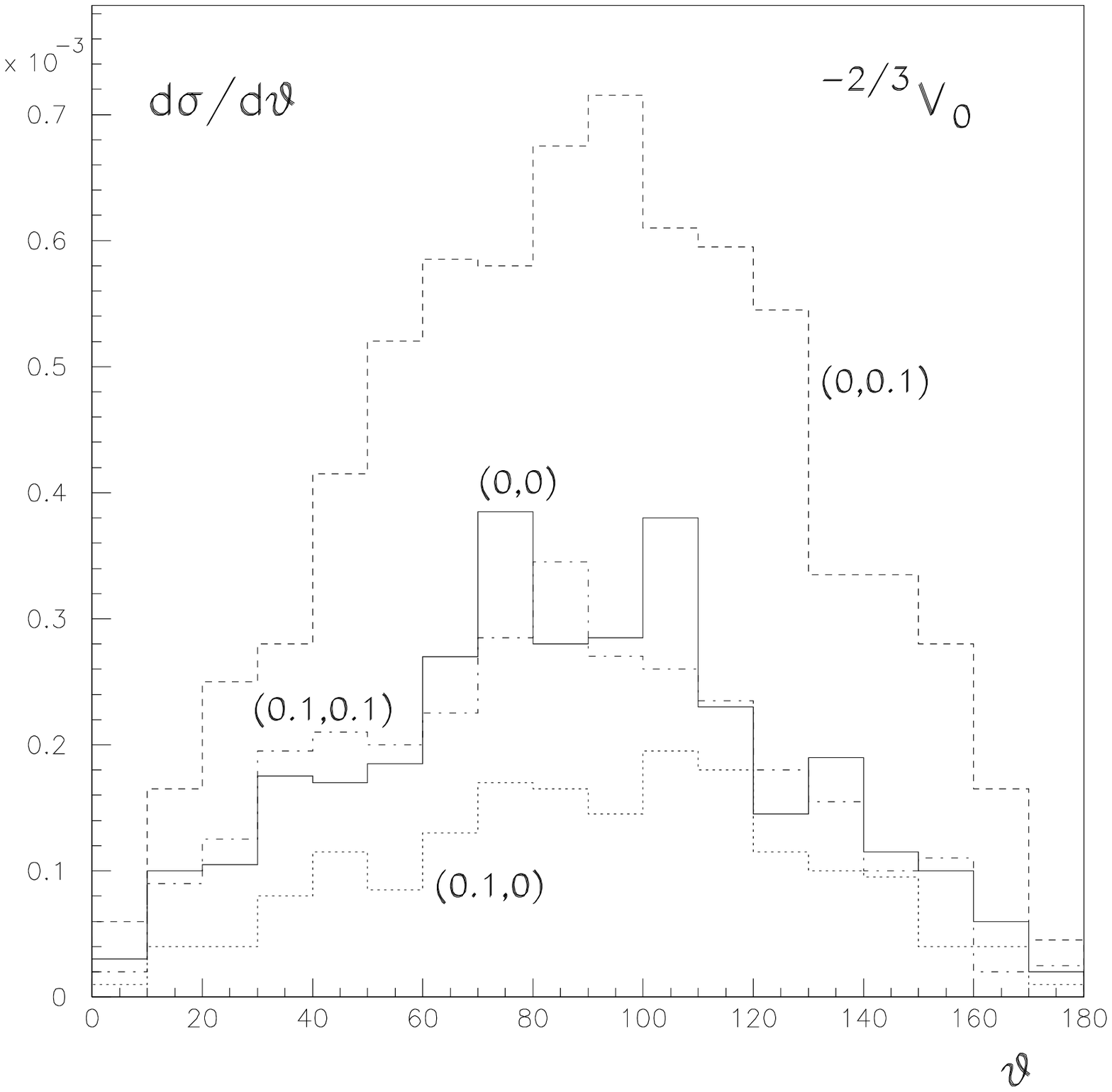}
}
\put(122,0){{\rm (b)}}
\end{picture}
\caption{\it Angular dependence of the production cross sections for
  (a) ${}^{-4/3}S_1$ and (b) ${}^{-2/3}V_0$ for various Yukawa
  couplings $(g_L,g_R)$ in units of $e$ ($M=300$ GeV,
  $\protect\sqrt{s} = 800$ GeV, ${\cal L} = 50$fb$^{-1}$, cuts
  included).}
\label{figsigang}
\end{figure}

Assuming that the masses are known and the Yukawa couplings not too
large, different species of leptoquarks can be distinguished already by
their total production cross section (see Fig.\ \ref{figsigtot} and
Table \ref{tabprop}). The angular distribution gives an additional
handle on the spin and the relative size of the couplings to gauge
bosons and fermions. For most of the leptoquarks in Table \ref{tabprop}
the $s/t$-channel interference is destructive reducing the cross section
in the central region. By contrast, the pure $t$-channel contribution
increases the cross section in the forward region. However, the Yukawa
couplings have to be relatively large to make this an observable effect.
In Fig.\ \ref{figsigang} the differential cross sections\footnote{Here,
  charge identification of the decay leptons is essential in order to
  remove the sign ambiguity of $\cos\theta$.  Experimentally this is
  possible for searches I and II even for very energetic
  electrons/positrons.} $d\sigma / d\theta$ including all cuts,
beamstrahlung and ISR are shown for ${}^{-4/3}S_1$ and ${}^{-2/3}V_0$
which have comparable production cross sections (see Table
\ref{tabprop}).  From the angular distribution it should not be
difficult to distinguish scalars from vectors, but the shape of the
distributions does not reflect the presence and size of Yukawa couplings
as clearly as one would wish.

\begin{figure}[htb] 
\unitlength 1mm
\begin{picture}(160,85)
\put(-3,-20){
\epsfxsize=9cm
\epsfysize=13cm
\epsfbox{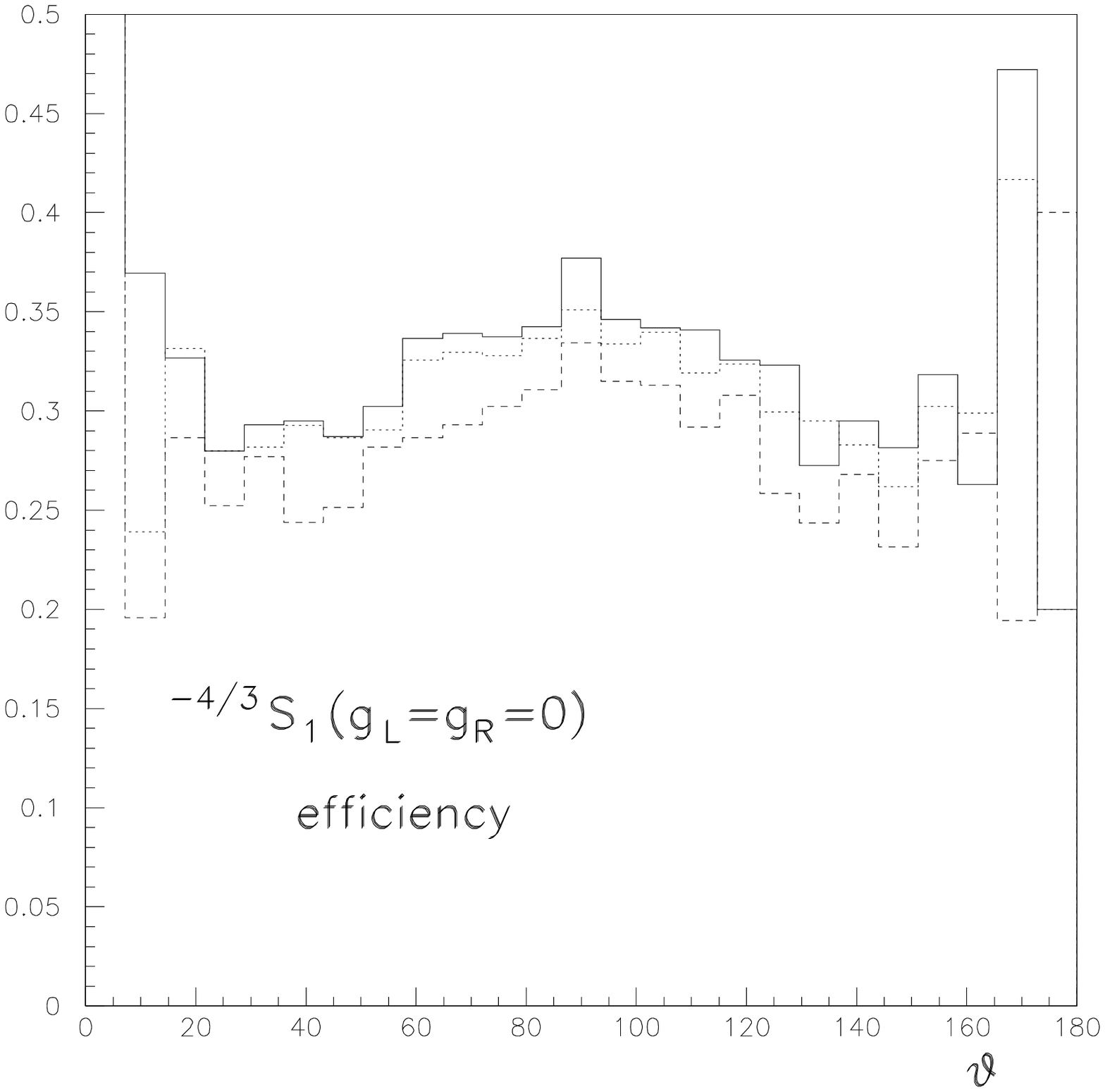}
}
\put(41,0){{\rm (a)}}
\put(78,-20){
\epsfxsize=9cm
\epsfysize=13cm
\epsfbox{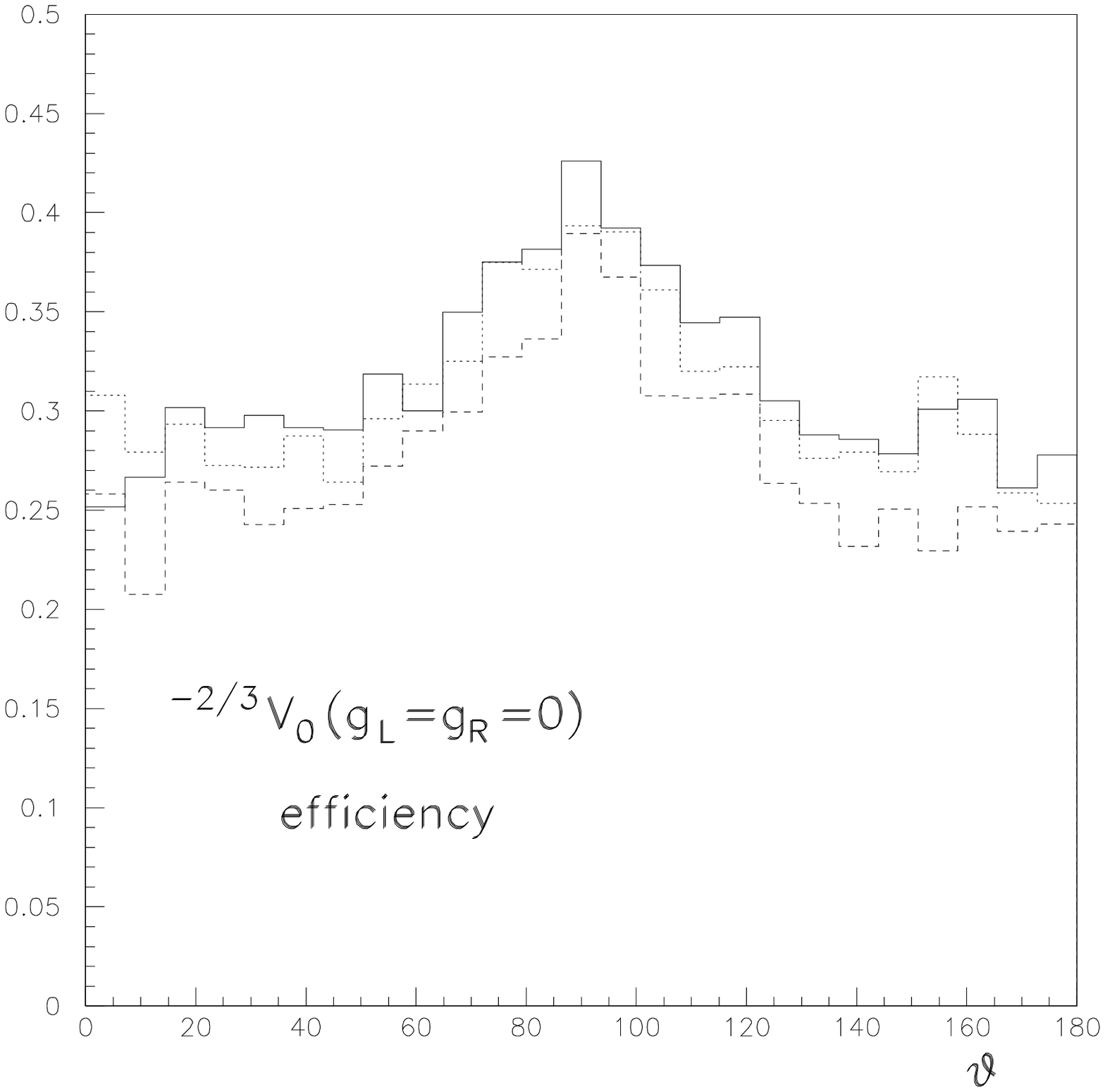}
}
\put(122,0){{\rm (b)}}
\end{picture}
\caption{\it Angular dependence of the detection efficiency for a
  scalar (a) and a vector (b) leptoquark in channel I. The full
  histogram refers to the 1 TeV detector, the dashed one to a
  LEP/SLD-type detector. The dotted histogram is obtained with the
  detector simulation of Ref.\ \protect\cite{simdet}.}
\label{figeffang}
\end{figure}

Another difficulty follows from the fact that, in general, the
detection efficiencies have an angular dependence which distorts the
observable distributions, in particular in the forward and backward
regions.  Figure \ref{figeffang} illustrates the angular dependence of
the detection efficiencies. For scalars (Fig.\ \ref{figeffang}a) the
cross section vanishes in the forward and backward region, leading to
large statistical fluctuations. Hence, detector effects show up mainly
in the normalization of the cross sections. However, for vector
leptoquarks (Fig.\ \ref{figeffang}b), the efficiency decreases by
almost 30\,\% at small and large angles. This demands a precise
unfolding of the efficiency, before one can hope to probe the Yukawa
couplings.

In order to evaluate the prospects for determining the couplings of
scalars more quantitatively, we have fitted the measured angular
distribution to the differential cross section
$d\hat{\sigma}/d\cos\theta$ given in Eq.\ (\ref{sigdiff}) with the
normalization as a free parameter. Consequently, one can only determine
ratios of the effective coupling parameters $g_1$, $g_2$, and $g_3$,
defined in Eqs.\ (\ref{coupg1} - \ref{coupg3}). Taking $M = 200$ GeV,
$\sqrt{s} = 500$ GeV and ${\cal L} = 20$fb$^{-1}$, it turns out that for
$|g_2/g_1| > 0.5$ and $g_3/g_1 > 0.15$ the ratios $g_2/g_1$ and
$g_3/g_1$ can be determined to 10\,\% accuracy. The above ranges
correspond to Yukawa couplings $\lambda_{L,R}$ between 0.2 and 0.6 (note
that the relation between the $g_i$ and $\lambda_{L,R}$ depends on the
leptoquark quantum numbers). This result does not change strongly with
$s$ and $M$. For $\lambda_{L,R} < 0.1$, the range allowed by low-energy
data (see Table \ref{tabprop}), it seems to be difficult to probe the
Yukawa couplings by investigating the angular distributions.

\section{Conclusions}

Linear $e^+e^-$ colliders provide unique tools to search for leptoquarks
independently of the size of their Yukawa couplings to lepton-quark
pairs.  The relatively clean environment of $e^+e^-$ annihilation allows
to reconstruct leptoquark masses from the decay products also in the
presence of beamstrahlung, QED initial state radiation, and
hadronization effects. The discovery limits depend on the quantum
numbers of the particular leptoquark and are estimated to reach $M =
(0.9$ to $0.95) \sqrt{s}/2$ in the most favorable cases.

The masses can be reconstructed with a precision of 0.5 to 0.8 GeV
assuming the detector performance expected for a dedicated 1 TeV
detector and not accounting for systematic mass shifts. Past experience
at LEP has shown that for the analysis of real data better measuring
accuracies can be achieved relative to early simulations.  Measurements
of the total cross sections in combination with an analysis of the
angular distributions will allow to clearly distinguish scalar from
vector leptoquarks. Yukawa couplings can be determined probably only if
they are of a strength comparable to the electromagnetic coupling.

In summary, searches for leptoquarks at $e^+e^-$ linear colliders are
complementary to $pp$ collisions which also probe the existence of
these novel states independently of the unknown Yukawa couplings, but
with less power in distinguishing states with different quantum
numbers. On the other hand, Yukawa couplings can be probed by
searching for virtual leptoquark effects in $e^+e^- \rightarrow
q\bar{q}$ and $pp \rightarrow e^+e^-$, and most efficiently in $ep$
collisions.


\end{document}